\documentclass[a4paper,fleqn]{cas-dc}

\usepackage[numbers]{natbib}

\usepackage{subcaption}

\usepackage{enumitem}

\usepackage{siunitx}

\usepackage{array}
\usepackage{booktabs}
\usepackage[table,xcdraw]{xcolor}

\PassOptionsToPackage{normalem}{ulem}
\usepackage{ulem}

\usepackage{float}

\def\tsc#1{\csdef{#1}{\textsc{\lowercase{#1}}\xspace}}
\tsc{WGM}
\tsc{QE}

\newcommand{\beq}{\begin{equation}}
\newcommand{\beql}[1]{\begin{equation}\label{#1}}
\newcommand{\eeq}{\end{equation}}
\newcommand{\bea}{\begin{eqnarray}}
\newcommand{\eea}{\end{eqnarray}}
\newcommand{\bean}{\begin{eqnarray*}}
\newcommand{\eean}{\end{eqnarray*}}

\newcommand{\mbf}[1]{\mathbf{#1}}

\newcommand{\ignore}[1]{}

\newcommand{\bF}{\mbf{F}}

\newcommand{\bK}{\mbf{K}}

\newcommand{\bU}{\mbf{U}}

\begin{document}
\let\WriteBookmarks\relax
\def\floatpagepagefraction{1}
\def\textpagefraction{0.2}

\shorttitle{SDF-based post-processing for SIMP topology optimization}    

\shortauthors{O. Jezek et~al.}  

\title [mode = title]{Smooth geometry extraction from SIMP topology optimization: Signed distance function approach with volume preservation}  

\tnotemark[1] 

\tnotetext[1]{Some note}

%

\author[1,2]{Ondřej Ježek}[orcid=0000-0002-5380-442X]
\ead{jezek@it.cas.cz}

\author[1]{Ján Kopačka}[orcid=0000-0002-2975-8347]
\ead{kopacka@it.cas.cz}

\author[1]{Martin Isoz}[orcid=0000-0002-5862-2561]
\ead{isozm@it.cas.cz}

\author[1]{Dušan Gabriel}[orcid=0000-0002-7691-2191]
\ead{gabriel@it.cas.cz}

\author[3]{Martin Šotola}[orcid=0000-0001-9938-4929]
\ead{sotola@vsb.cz}

\author[3]{Pavel Maršálek}[orcid=0000-0002-4826-0755]
\ead{marsalek@vsb.cz}

\author[3]{David Rybanský}[orcid=0000-0002-5201-9606]
\ead{rybansky@vsb.cz}

\author[3]{Radim Halama}[orcid=0000-0002-3546-4660]
\ead{rybansky@vsb.cz}

\affiliation[1]{organization={Institute of Thermomechanics, Czech Academy of Sciences},
            addressline={Dolejškova 1402/5}, 
            postcode={182 00, Praha 8}, 
            state={Czech Republic}}

\affiliation[2]{organization={Faculty of Mechanical Engineering, Czech Technical University in Prague},
            addressline={Technická 4}, 
            postcode={160 00, Praha 6}, 
            state={Czech Republic}}

\affiliation[3]{organization={Faculty of Mechanical Engineering, VSB – Technical University of Ostrava}, 
            addressline={17. listopadu 2172/15}, 
            postcode={708 00}, 
            city={Ostrava}, 
            state={Czech Republic}}










\begin{abstract}
This paper presents a novel post-processing methodology for extracting high-quality geometries from density-based topology optimization results. Current post-processing approaches often struggle to simultaneously achieve smooth boundaries, preserve volume fraction, and maintain topological features. We propose a robust method based on a signed distance function (SDF) that addresses these challenges through a two-stage process: first, an SDF representation of density isocontours is constructed, which is followed by geometry refinement using radial basis functions (RBFs).
The method generates smooth boundary representations that appear to originate from much finer discretizations while maintaining the computational efficiency of coarse mesh optimization. Through comprehensive validation, our approach demonstrates a 18\% reduction in maximum equivalent stress values compared to conventional methods, achieved through continuous geometric transitions at boundaries. The resulting implicit boundary representation facilitates seamless export to standard manufacturing formats without intermediate reconstruction steps, providing a robust foundation for practical engineering applications where high-quality geometric representations are essential.
\end{abstract}





\maketitle

\section{Introduction}
\noindent With the increasing availability of additive manufacturing and other modern manufacturing methods, topology optimization (TO) has become an invaluable tool for structural design, enabling engineers to discover novel geometries. This approach to computational design determines the optimal material distribution within a design space to achieve specified performance objectives while satisfying given constraints~\cite{Tsavdaridis2015,Andreassenkey2011}. The field has rapidly evolved from academic research into practical engineering applications across aerospace~\cite{Zhu2016}, automotive~\cite{Tang2014}, and biomedical industries~\cite{Wu2021}. Among various TO methods, density-based approaches like the Solid Isotropic Material with Penalization (SIMP) method have become predominant due to their mathematical simplicity and robust convergence properties~\cite{Rozvany2000, LI2022}. However, the raw results from density-based TO typically exhibit two significant limitations: jagged boundaries due to the underlying finite element discretization and regions of intermediate density values that do not represent physically meaningful material states~\cite{VanDijk2013, subedi2020}.

These limitations present significant challenges in translating optimized designs into practical applications. Consequently, the development of effective post-processing methods has become a critical area of research~\cite{subedi2020}. Several key factors must be considered in the development of such methods:

\begin{enumerate}
\item \textbf{Geometric fidelity}: The post-processed geometry must preserve the fundamental structural features and topological characteristics of the optimized solution while eliminating artificial artifacts.
\item \textbf{Volume preservation}:  The specified material volume fraction from the original optimization must be maintained to ensure design feasibility and performance requirements remain satisfied.
\item \textbf{Boundary quality}: The extracted geometry should exhibit smooth boundaries to reduce stress concentrations and improve manufacturability.
\item \textbf{Mesh independence}: The geometry extraction procedure should be applicable across various discretization schemes used in topology optimization, including regular and irregular meshes, elements of different polynomial orders, and geometrically diverse element types.
\item \textbf{Multi-platform compatibility}: The resulting geometry must be compatible with downstream engineering processes, such as computer-aided design (CAD) systems, finite element analysis (FEA) software, or manufacturing processes.
\end{enumerate}

Recent research efforts have explored various approaches to address the challenges arising from considering the above-listed factors. First, image processing techniques employ thresholding operations followed by boundary smoothing using control points and spline interpolation~\cite{Lin2000}. While computationally efficient, these methods often struggle to simultaneously preserve volume fractions and critical geometric features. Next, model reconstruction approaches attempt to fit geometric primitives or spline surfaces to the optimized topology~\cite{Tang2001}, offering improved boundary smoothness but frequently introducing significant computational complexity and potential loss of topological features.

More recent developments have shown promising advances in addressing multiple requirements simultaneously. Li et al.~\cite{Lii2021} introduced a boundary density evolution method that combines density-based optimization with level-set smoothing to produce manufacturable designs without penalty factors. While the method generates smooth boundaries suitable for 3D printing, its implementation is limited by its computational complexity and restriction to structured meshes.

In parallel, iso-density contour methods have demonstrated significant potential for balanced post-processing solutions. Li et al.~\cite{LI2022} developed a novel lookup table-based smoothing method for topology optimization results that achieves both computational efficiency and volumetric preservation. Their method shows particular strength in feature preservation and processing speed, though additional steps including internal surface removal and remeshing algorithms are required for three-dimensional structures.

Despite these advances, several limitations persist in current post-processing approaches. First, many methods struggle to simultaneously achieve boundary smoothness and precise volume preservation, particularly for complex geometries. Second, the computational efficiency of existing approaches often deteriorates with increasing mesh refinement or geometric complexity. Third, while some methods work well for regular meshes, their performance on irregular discretizations or higher-order elements remains inconsistent. Finally, the generation of CAD-compatible geometries, especially for three-dimensional problems, continues to pose significant challenges~\cite{subedi2020}.

This paper builds upon principles established in~\cite{VanDijk2013} and extends the work of~\cite{Swierstra2017}, which introduced an effective method for extracting geometries from SIMP-based results while preserving critical geometric features, maintaining precise volume fractions, and generating smooth boundaries through subsequent level-set shape optimization. This work extends this methodology by developing a robust framework capable of handling various discretization schemes, including irregular meshes, higher-order elements, and geometrically diverse element types. This enhanced capability enables the transition of geometry extraction methods from purely academic applications to practical industrial problems.

The paper is organized as follows: Section \ref{sect:SIMP} first provides a brief overview of the SIMP method for topology optimization. Section \ref{sect:methodology} introduces the core methodology for geometry extraction. Section \ref{sect:detailed_methodology} presents a detailed examination of the geometry extraction methodology, including implementation considerations and numerical aspects. Section \ref{sect:num_tests} is dedicated to comprehensive testing and validation of the proposed method's effectiveness through various case studies. Finally, conclusions and recommendations for future work are presented in Section \ref{sect:conclusion}. Appendix \ref{app:distance_function} provides a detailed implementation-oriented description of the distance function construction methodology, complementing the theoretical foundation presented in Section \ref{sect:detailed_methodology}.

\section{SIMP method for structural topology optimization}\label{sect:SIMP}
\noindent For the sake of completeness of this work, we briefly outline the SIMP (Solid Isotropic Material with Penalization) method here. This method is employed in this study for structural topology optimization~\cite{Bendsoe1989,Zhou1991,Mlejnek1992}. 

\subsection{Problem formulation}\label{sub:SIMP-problem_formulation}
\noindent In the implementation of the SIMP method~\cite{Andreassenkey2011}, the design domain is discretized into a finite element mesh, where the density of each element can take any value from 0 (void) to 1 (solid). The objective is to minimize structural compliance (maximize stiffness) while satisfying given constraints. The discrete formulation is:
\begin{align}
  \text{Minimize}: \quad &C  = \bF^\mathsf{T} \bU \notag\\
    \text{Subject to}: \quad &V^*-\sum_{e=1}^N V_e \rho_e=0 \label{eqn:comp_mini}\\
                                   &\bK \bU = \bF \notag \\
    &0\leq \rho_e \leq 1, \notag
\end{align}
\noindent where \(C\) represents the compliance, \( \bK\) is the global stiffness matrix, \( \bU\) is the displacement vector, \(V^*\) is the target structural volume, and \(V_e\) is the volume of the \(e\)-th element. The design variable \( \rho_e\), ranging from 0 to 1, indicates the material density at each element. The equilibrium constraint, \(\bK \bU = \bF\), ensures that the structural deformations under the applied forces are accurately accounted for, where \(\bF\) is the vector of external forces acting on the structure. 

In the modified SIMP method~\cite{Andreassenkey2011}, the material model for each element is defined as a function of its density:
\begin{equation}
  E_e = E_{min} + \rho_e^p(E_0-E_{min})\,, \label{eqn:mod_simp}
\end{equation}
\noindent where \(E_0\) is the Young's modulus of the solid material, and \(E_{\text{min}}\) is a small value assigned to avoid numerical instabilities in regions with zero density. The exponent \(p\), known as the penalization parameter, is typically set to 3~\cite{Bendsoe1999}. This parameter penalizes density values between 0 and 1, thereby preventing intermediate densities from occurring in the final optimized structure and promoting a clear separation between solid (material) and void (no material) regions.

\subsection{Solution method}\label{sub:SIMP-solution_method}
\noindent A common method to solve topology optimization problems using the SIMP approach is the Optimality Criteria (OC) method~\cite{Sigmund2001}. This iterative method adjusts the density of each element to minimize structural compliance while respecting the volume constraint. The update scheme for the density of each element can be described as follows:
\begin{equation}
\rho^{\text{new}}_e = 
\begin{cases}
\max(0, \rho_e - m) & \text{if } \quad \rho_e B^\eta_e \leq \max(0, \rho_e - m) \\
\min(1, \rho_e + m) & \text{if } \quad \rho_e B^\eta_e \geq \min(1, \rho_e + m) \\
\rho_e B^\eta_e & \text{otherwise},
\end{cases}
\end{equation}
\noindent where $m$ is the maximum allowed predefined density change per iteration, which limits the step size to maintain numerical stability. The exponent $\eta$ acts as a penalization or damping factor to further ensure smooth and stable updates. The sensitivity measure $B_e$, which determines both the direction and magnitude of the density update, is defined as:
\begin{equation}
    B_e = -\frac{\partial C}{\partial \rho_e} / \lambda \frac{\partial V}{\partial \rho_e},
\end{equation}
\noindent where \(\lambda\) is the Lagrange multiplier, determined iteratively to satisfy the volume constraint. The sensitivity of the compliance \(C\) with respect to the density \(\rho_e\) of an element is calculated by:
\begin{equation}
  \frac{\partial C}{\partial \rho_e} = -p \rho_e^{p-1} (E_0 - E_{\text{min}}) \boldsymbol{u}_e^\mathsf{T} \boldsymbol{k}_0 \boldsymbol{u}_e\,,
\end{equation}
\noindent where \(\boldsymbol{u}_e\) represents the displacement vector of the element, and \(\boldsymbol{k}_0\) is the elemental stiffness matrix for a fully solid material with unit Young’s modulus.

\subsection{Density filter for numerical stability}\label{sub:SIMP-filter}
\noindent To ensure computational stability and to eliminate the well-known checkerboard pattern, which leads to non-physical material distributions, density filtering techniques are employed~\cite{Sigmund2007}. Among these, the density filter approach~\cite{Tortorelli2001,Bourdin2001} is widely used to address mesh dependency and eliminate numerical instabilities by smoothing the field of design variables. This filtering process averages the densities within a local neighborhood, with the operation expressed as:
\begin{equation}
    \tilde{\rho}_e = \frac{\sum_{i \in N_e} H_{ei}\rho_i}{\sum_{i \in N_e} H_{ei}}\,,
\end{equation}
\noindent where \(H_{ei}\) is a weighting factor that depends on the distance between element \(e\) and its neighbors \(N_e\). The filtered density \(\tilde{\rho}_e\) represents a physically meaningful density, addressing issues of mesh dependency and numerical instabilities. Throughout this paper, we refer to the field of filtered densities simply as \(\rho\).

\section{Methodology for geometry extraction}\label{sect:methodology}
\begin{figure*}[h!]
  \centering
  \includegraphics[clip,width=0.95\textwidth]{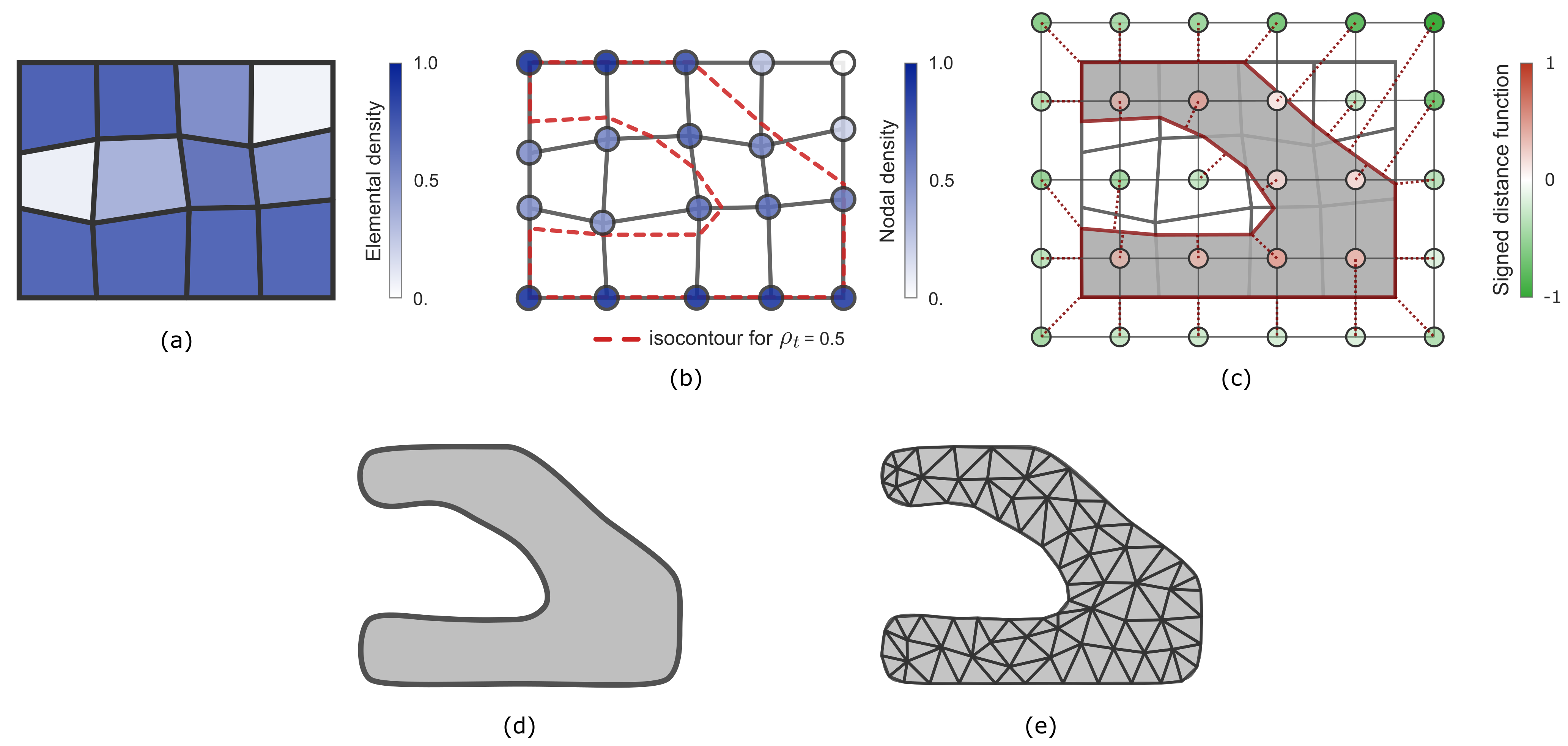}
  \caption{An illustration of the procedure for extracting a smooth geometry from topology optimization results, demonstrated on a 2D beam example. (a) Input data: visualization of raw topology optimization result from the SIMP method, represented as an element-wise density field. (b) Mapping elemental densities to nodal values and extracting the geometry using an isocontour technique. (c) Constructing an SDF on a regular Cartesian grid to mathematically define the boundary. (d) Refining the geometry by smoothing the SDF with radial basis functions, resulting in an accurate and smooth boundary. (e) Discretizing the smoothed geometry into a high-quality mesh for subsequent applications.}
  \label{fig:Postup_extrakce}
\end{figure*}

\noindent To facilitate navigation through the technical section of this work, we first provide a brief introduction of the newly proposed methodology for extracting geometry from raw topology optimization results. An extensive description of the methodology is given in Section~\ref{sect:detailed_methodology}. The main steps of the methodology are illustrated in Figure~\ref{fig:Postup_extrakce}. With input of the raw topology optimization results (Figure~\ref{fig:Postup_extrakce}a), they comprise:
\begin{enumerate}
  \item \textbf{Mapping elemental densities to nodes} (Figure~\ref{fig:Postup_extrakce}b): Elemental densities from the optimization results are mapped to the nodes of the original unstructured mesh, establishing the nodal values essential for subsequent geometry definition. This mapping is detailed in Section~\ref{sub:nodal-dense}.
  \item \textbf{Geometry definition through isocontour technique} (Figure~\ref{fig:Postup_extrakce}b): Nodal density values are interpolated using finite element method (FEM) shape functions within the element space. The boundary emerges naturally as an isocontour where this interpolated density field equals a predefined threshold value.
  \item \textbf{Signed distance function (SDF) construction} (Figure~\ref{fig:Postup_extrakce}c): The density isocontours are then remapped using an SDF defined on a regular Cartesian grid, which provides a mathematical foundation for subsequent geometric refinement, see Section~\ref{sub:SDF} for details.
  \item \textbf{Smoothing using radial basis functions (RBFs)} (Figure~\ref{fig:Postup_extrakce}d): The SDF is refined using RBFs to ensure boundary smoothness and geometric accuracy. This step is detailed in Section~\ref{sub:Geometry_smoothing}.
  \item \textbf{Discretization for engineering applications} (Figure~\ref{fig:Postup_extrakce}e): The smoothed geometry is discretized into a high-quality mesh suitable for subsequent engineering applications, including finite element analysis, manufacturing processes, or further geometric refinements. This process is covered in Section~\ref{sub:Discretization}.
\end{enumerate}

\section{Detailed methodology}\label{sect:detailed_methodology}
\noindent This chapter provides a detailed methodology for transforming raw topology optimization results from density-based methods, particularly SIMP, into refined geometric structures.

\subsection{Density field processing} \label{sub:nodal-dense}
\noindent The SIMP method produces a design defined by elements and their densities. Since these densities are element-wise constant, they are unsuitable for further processing. Thus, it is necessary to generate a nodal density field from them. One suitable approach is to map the elemental densities onto the nodes of the same mesh~\cite{Swierstra2017}.

Several techniques exist for computing nodal density values. Basic approaches employ simple averaging of densities from elements sharing a node, while more sophisticated methods utilize interpolation or approximation techniques, such as the Superconvergent Patch Recovery method commonly used in finite element analysis~\cite{Zienkiewicz2013}.

In this paper, we have chosen an efficient linear approximation method, which fits the elemental densities of neighboring elements to each node of the same mesh using a linear polynomial in the least squares sense. This approach was selected for its ability to preserve the geometry of the optimized object, particularly when applied to general (irregular) meshes. Such preservation is essential, as simpler averaging methods can result in significant geometric distortion.

The linear polynomial used to approximate the density at each node takes the form:
\begin{equation}
\rho(\boldsymbol{x}) =  \boldsymbol{a}^\mathsf{T}(1,\boldsymbol{x})^\mathsf{T}, \label{eqn:lin_inter_lsq}
\end{equation}
where $\boldsymbol{x} \in \mathbb{R}^3$ represents a point in space and $\boldsymbol{a} \in \mathbb{R}^4$ contains the polynomial coefficients.
To determine these unknown coefficients, we formulate a least squares problem:
\begin{equation}
\mathbf{X} \mathbf{a} = \boldsymbol{\rho}\,, \label{eqn:Least_squares}
\end{equation}
where the design matrix $\mathbf{X} \in \mathbb{R}^{m \times 4}$ contains coordinates of geometric centers of elements sharing the given node, with $m$ being the number of elements connected to that node, and vector $\boldsymbol{\rho} \in \mathbb{R}^{m}$ contains the elemental densities.

The system of linear equations for the least squares fitting is solved to determine the coefficients $\boldsymbol{a}$:
\begin{equation}
  \mathbf{X}^\mathsf{T} \mathbf{X} \mathbf{a} = \mathbf{X}^\mathsf{T} \boldsymbol{\rho}\,.
\end{equation}
Once the coefficients are determined for a particular node, the corresponding nodal density can be computed using equation \eqref{eqn:lin_inter_lsq} by substituting the node's coordinates $\boldsymbol{x}$. Since each nodal value can be calculated independently of others, this approach is well-suited for parallel computation, enabling efficient processing of large meshes. 

The result is a set of discrete density values assigned to mesh nodes that effectively preserves the key geometric features of the optimized structure.

\subsection{Geometry representation using signed distance function} \label{sub:SDF}
\noindent The SDF is a widely used mathematical tool in computational geometry, computer graphics, and structural optimization~\cite{Stanley2004}. It characterizes geometric shapes by assigning to each point in space a scalar value equal to its minimum distance from a boundary surface, with the sign convention indicating the point's position relative to the boundary (positive inside, negative outside). SDFs are particularly valuable in topology optimization applications due to their ability to efficiently represent complex geometries. The definition of an SDF $\phi(\boldsymbol{x})$ for a point $\boldsymbol{x} \in \mathbb{R}^3$ in space can be expressed as follows:
\begin{equation} \label{eq:SDF_form}
\phi(\boldsymbol{x}) =
\begin{cases} 
d(\boldsymbol{x}, \partial \Omega) & \text{if } \boldsymbol{x} \in \Omega \\
0 & \text{if } \boldsymbol{x} \in \partial \Omega \\
-d(\boldsymbol{x}, \partial \Omega) & \text{if } \boldsymbol{x} \in \Omega^c,
\end{cases}
\end{equation} 
\noindent where $d(\boldsymbol{x}, \partial \Omega)$ represents the Euclidean distance between the point $\boldsymbol{x}$ and the boundary $\partial \Omega$ of the domain $\Omega$, and $\Omega^c$ is the complement of $\Omega$.

The particular form of the function~\eqref{eq:SDF_form} can in principle be defined in both continuous and discrete forms. This work adopts the discrete representation due to its practical advantages for topology optimization post-processing. 
Specifically, discrete SDFs provide reliable representation of complex geometrical features in topology optimization. This approach preserves geometric detail while offering potential for adaptive enhancement~\cite{koschier2017hp}.

The computation of the discrete form of the SDF is typically divided into two separate steps: distance calculation and sign assignment. The distance calculation determines the shortest Euclidean distance between each evaluation point and the boundary surface. Sign assignment identifies whether the point is inside or outside the object, often by evaluating the orientation of the normal vector at the boundary~\cite{Curless1996} or performing a point-in-polygon test~\cite{hughes2014computer}. Separation of the distance calculation and sign assignment simplifies implementation by treating them as independent operations, facilitates the validation of intermediate results, and improves overall code efficiency.

In the proposed methodology, we transfer the object boundary represented by a density isosurface on an irregular mesh to a signed distance function on a regular and optionally finer Cartesian grid. This transformation facilitates the subsequent smoothing of the geometry using RBFs, as detailed in Section \ref{sub:Geometry_smoothing}.

\subsubsection{Construction of distance function} \label{subsub:distance_function_formulation}
\noindent The fundamental concept of constructing a signed distance function involves calculating the shortest distance between nodes of a regular Cartesian grid and the material domain boundary.
This boundary is defined as the isocontour where the interpolated density $\rho(\boldsymbol{\xi})$ equals the threshold value $\rho_t$. The mathematical formulation for this distance computation can be expressed through an optimization problem:

\begin{equation} \label{eq:dist_mini}
\begin{cases}
\text{find} & \quad \boldsymbol{\xi} = \arg \min d(\boldsymbol{\xi}) \\
\text{subjected to} & \quad \rho(\boldsymbol{\xi}) = \rho_t \\
& \quad -1 \leq \xi_i \leq +1, \\
\end{cases}
\end{equation}
where the constraint $-1 \leq \xi_i \leq +1$ ensures that the projection remains within the element domain.\\

\begin{figure}[h!]
    \centering
    \includegraphics[width=0.35\textwidth]{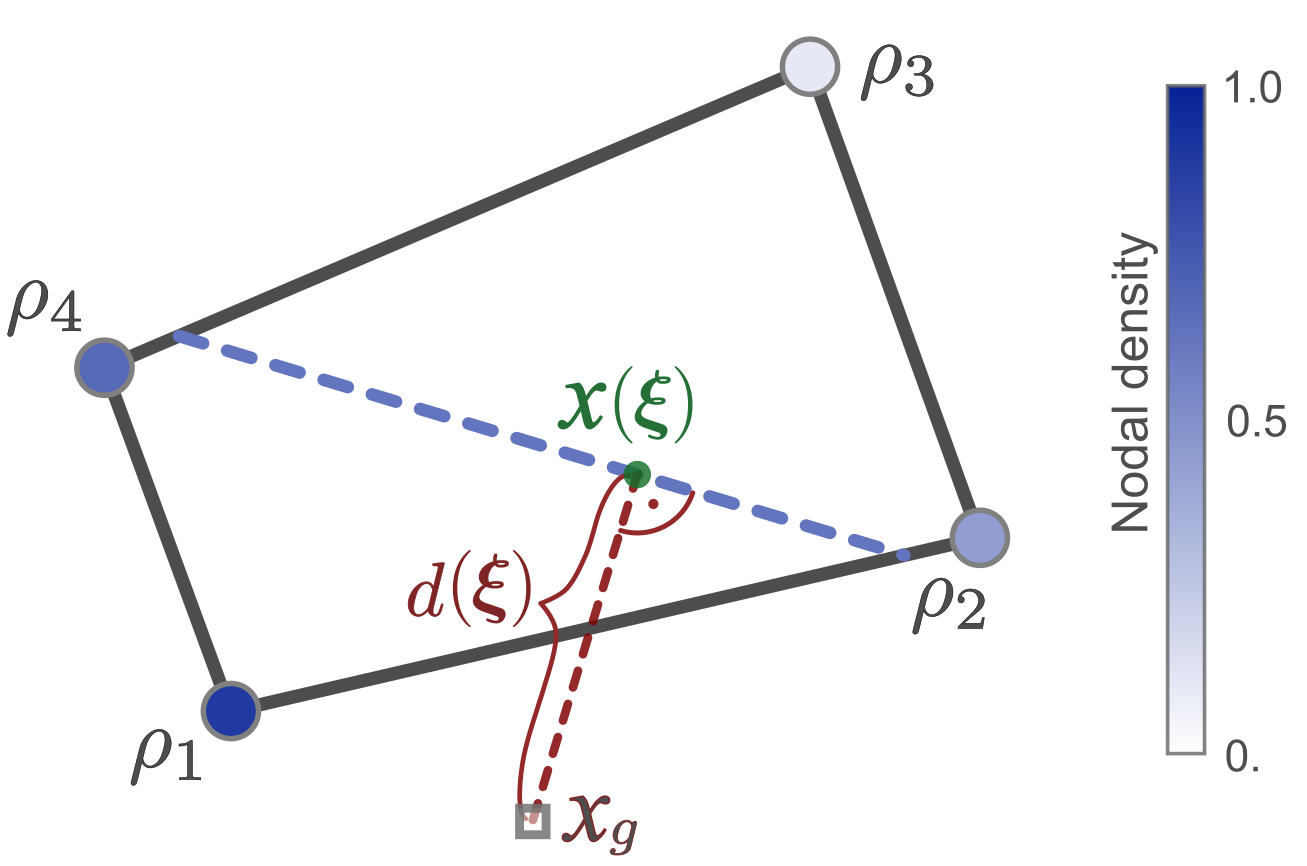}
    \caption{Illustration of distance calculation from a Cartesian grid point $\boldsymbol{x}_g$ to the density isocontour $\rho = \rho_t$ within a finite element. The shortest distance is represented by the perpendicular projection onto the isocontour.}
    \label{fig:Distance_field_element}
\end{figure}

\noindent The distance function is defined as:
\begin{equation}
d(\boldsymbol{\xi})= \|\boldsymbol{x}_{g}-\boldsymbol{x}(\boldsymbol{\xi})\|, \label{eq:rhoxi}
\end{equation}
where $\boldsymbol{\xi}$ represents the local coordinates within an element where we seek the projection point, and $\boldsymbol{x}_g$ represents the position of the grid node for which we calculate the distance (see Figure~\ref{fig:Distance_field_element}).
The interpolation of nodal densities and computation of local coordinates for points projected onto the isocontour can be performed using standard finite element shape functions:
\begin{align}
  \rho(\boldsymbol{\xi})=\sum_{a=1}^{n_{\mathrm{en}}}N_a(\boldsymbol{\xi}) \rho_a, \label{eq:int_density} \\
  \boldsymbol{x}(\boldsymbol{\xi})=\sum_{a=1}^{n_{\mathrm{en}}}N_a(\boldsymbol{\xi}) \boldsymbol{x}_a. \label{eq:xxi_coord}
\end{align}
Here, $N_a$ are the element shape functions, while $\rho_a$ and $\boldsymbol{x}_a$ represent nodal density values and coordinates, respectively.

\begin{figure}[h!]
    \centering
    \includegraphics[width=0.35\textwidth]{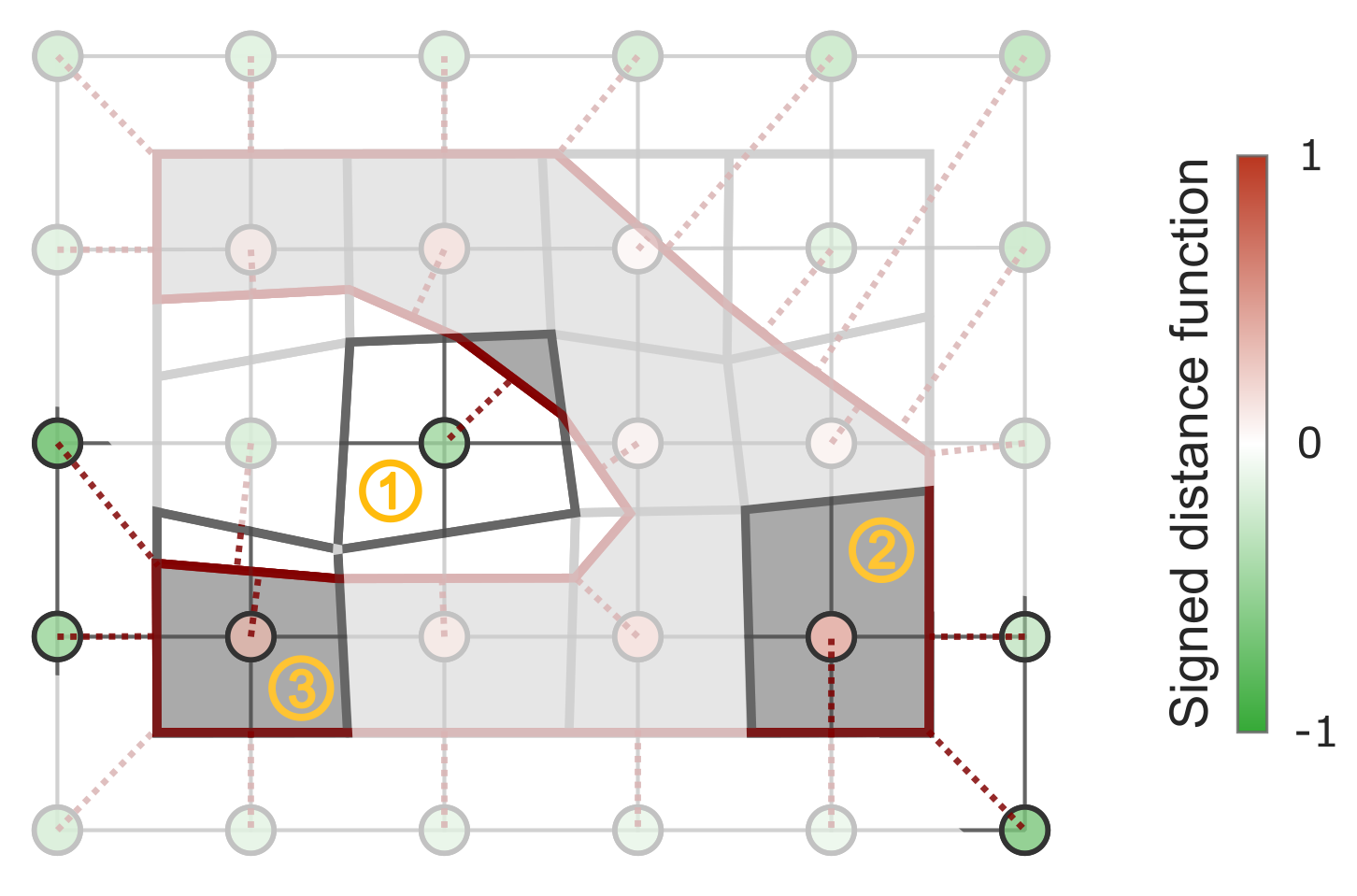}
    \caption{Visualization of different geometric configurations and corresponding projection scenarios in distance function construction. The Figure~shows three geometric configurations: an element with internal isocontour (1), a fully solid boundary element (2), and a transitional element with partially solid boundary face (3).}
    \label{fig:Druhy_projekce}
\end{figure}

While equation \eqref{eq:dist_mini} provides a general formulation for distance calculation onto the isocontour, its solution encompasses more than just the commonly assumed perpendicular projection. The practical implementation must carefully address multiple possible geometric configurations. As illustrated in Figure~\ref{fig:Druhy_projekce}, we identify three distinct geometric configurations that require different implementations of the distance function:
\begin{enumerate}
  \item Elements containing an isocontour within the topology optimization domain (detailed in Section \ref{subsubapp:dist2iso})
  \item Fully solid elements at the domain boundary (detailed in Section \ref{subsubapp:dist2face})
  \item Transitional elements with partially solid faces at the domain boundary (detailed in Section \ref{subsubapp:dist2transface})
\end{enumerate}
For 2D problems, each element type involves two possible projection types: projection onto an edge or onto a vertex. In 3D applications, a third projection type emerges - projection onto a surface - resulting in nine distinct projection scenarios that must be handled appropriately.

Details of the implementation of these processes, including the selection of the optimal grid resolution and boundary detection algorithms, are thoroughly described in Appendix~\ref{app:distance_function}.

\subsubsection{Sign assignment}\label{subsub:sign}
\noindent To assign the required signs to the distance function values, we have developed a density-based evaluation approach that compares the local density $\rho(\boldsymbol{\xi})$ at each point $\boldsymbol{x}(\boldsymbol{\xi})$ against a threshold density $\rho_t$. This evaluation requires determining whether a given node lies within an element of the original mesh and calculating its corresponding density value.

For each regular grid node $\boldsymbol{x}_g$ we first identifying which element of the original mesh contains the node, and then calculate its local coordinates $\boldsymbol{\xi}$ within that element by solving this equation:
\begin{equation}
\|\boldsymbol{x}_{g}-\boldsymbol{x}(\boldsymbol{\xi})\| = 0
\end{equation}
where $\boldsymbol{x}(\boldsymbol{\xi})$ is the point position computed by equation \eqref{eq:xxi_coord}.

Once the local coordinates are established, the density at this point can be computed using finite element shape functions $N_a(\boldsymbol{\xi})$ and nodal density values $\rho_a$. Based on the density field and the position of the point relative to the density isocontour, the sign can be determined.
\subsubsection{Implementation details}\label{subsub:sign_implement}
\noindent The practical implementation of sign assignment follows a systematic five-step process:
\begin{enumerate}
  \item \textbf{AABB generation:}
First, Axis-Aligned Bounding Boxes (AABBs)~\cite{Fujun2000, Chen2014} are created for all elements from the original mesh that contain at least one node with density higher than the threshold density $\rho_t$. This initial spatial partitioning provides an efficient framework for subsequent node classification.
\item \textbf{Node-element association:}
The algorithm iterates through each node of the regular grid to determine which AABBs contain it. This operation generates a data structure that maps each node to potential containing elements from the original mesh through their unique element IDs, significantly reducing the search space for subsequent operations.
\item \textbf{Element membership verification:}
For each identified node-element pair, the algorithm determines the local coordinates $\boldsymbol{\xi}$ by solving the equation $\|\boldsymbol{x}_{g}-\boldsymbol{x}(\boldsymbol{\xi})\| = 0$. This nonlinear system is solved through minimization:
\begin{equation} \label{eq:find_loccoords}
\boldsymbol{\xi} = \arg \min \|\boldsymbol{x}_{g}-\boldsymbol{x}(\boldsymbol{\xi})\|
\end{equation}
Node membership is confirmed when the computed local coordinates satisfy $-1 \leq \xi_i \leq 1$ for all coordinate components, establishing the precise element containing the node.
\item \textbf{Local density computation:}
Once the containing element is identified, the algorithm calculates the density at point $\boldsymbol{x}(\boldsymbol{\xi})$ using equation \eqref{eq:int_density}.
\item \textbf{Sign assignment:}
The final step assigns signs to nodes based on the computed density values:
\begin{itemize}
\item Positive sign (+): Assigned when $\rho(\boldsymbol{\xi}) \geq \rho_t$
\item Negative sign (-): Retained for nodes either outside all elements or where $\rho(\boldsymbol{\xi}) < \rho_t$
\end{itemize}
\end{enumerate}
This procedure ensures accurate and efficient sign determination while maintaining computational practicality for large-scale optimization problems.

\subsection{Geometry smoothing} \label{sub:Geometry_smoothing}
\noindent An important step in generating high-quality post-processed geometries is the refinement of the SDF. This section presents our approach in which we use RBFs to enhance the geometric representation while improving both the manufacturability and mechanical performance of the final design~\cite{Carr2001}.

\subsubsection{Radial basis functions}\label{subsub:RBF_theory}
\noindent RBFs provide a effective framework for transforming discrete geometric data into smooth continuous functions. The key advantage of RBFs lies in their ability to generate smooth transitions across the design domain while preserving essential geometric features~\cite{VanDijk2013}. These axisymmetric functions depend solely on the Euclidean distance between an evaluation point $\boldsymbol{x}$ and a center point $\boldsymbol{x_g}$, which are in our case nodes of the regular grid.

Based on extensive analysis of various RBF formulations~\cite{Wang2006}, we employ Gaussian functions for their $C^{\infty}$ continuity and proven effectiveness in representing optimized structural geometries. The Gaussian RBF is defined as:
\begin{equation}
R(r) = e^{-\left(\frac{r}{B}\right)^2}
\end{equation}
where $r = \|\boldsymbol{x} - \boldsymbol{x}_g\|$ is the Euclidean distance and $B > 0$ represents a scaling parameter. We set $B$ equal to the regular grid spacing $h$, ensuring an optimal RBF support size relative to the discretization and satisfying the partition of unity property~\cite{Wendland2004}.

\subsubsection{Smoothing implementation} \label{subsub:RBF_smoothing}

\noindent The smoothing process begins with a discrete dataset consisting of grid node coordinates $\{\boldsymbol{x}_i\}_{i=1}^n$ and their corresponding SDF values $\{\phi_i\}_{i=1}^n$, where $n$ represents the total number of grid nodes. We construct a smoothed SDF $\tilde{\phi}(\boldsymbol{x})$ through RBF interpolation that preserves essential geometric characteristics while eliminating discretization artifacts.

The RBF interpolation framework represents the smoothed function as a linear combination of basis functions centered at each grid node:
\begin{equation}
\tilde{\phi}(\boldsymbol{x}_g) = \sum_{i=1}^n s_i R(\|\boldsymbol{x}_g - \boldsymbol{x}_i\|)
\end{equation}
where $s_i$ are interpolation weights to be determined, and $R(\cdot)$ is the Gaussian RBF defined in the previous section. The interpolation weights are computed by enforcing the condition that the RBF interpolation exactly reproduces the given SDF values at all grid nodes:
\begin{equation}
\tilde{\phi}(\boldsymbol{x}_j) = \sum_{i=1}^n s_i R(\|\boldsymbol{x}_j - \boldsymbol{x}_i\|) = \phi_j, \quad \forall j \in \{1, ..., n\}
\end{equation}
This condition leads to a system of linear equations that can be expressed in matrix form:
\begin{equation}
\mathbf{A} \mathbf{s} = \boldsymbol{\phi}
\label{eq:linearRBFs}
\end{equation}
where $\mathbf{A} \in \mathbb{R}^{n \times n}$ is the RBF interaction matrix with elements $A_{ij} = R(\|\boldsymbol{x}_i - \boldsymbol{x}_j\|)$, $\mathbf{s} \in \mathbb{R}^n$ is the vector of unknown weights, and $\boldsymbol{\phi} \in \mathbb{R}^n$ contains the original SDF values. The weight vector is obtained by solving this linear system~\eqref{eq:linearRBFs}.\\

To ensure that the smoothed geometry maintains the same volume as the original design, a uniform shift parameter $c$ is applied to the interpolated SDF. This volume-preserving correction modifies the zero-level set position without altering the geometric smoothness achieved through RBF interpolation. The optimal value of $c$ is determined through an iterative process that adjusts the parameter until the zero-level set of the shifted function encloses a volume that matches the target volume within a prescribed tolerance.

For computational efficiency in large-scale applications, the implementation can benefit from localized RBF support. When RBF influence is restricted to a neighborhood around each center point, the interaction matrix $\mathbf{A}$ becomes sparse, enabling the use of efficient sparse matrix solvers. Additionally, spatial data structures such as K-D trees can be employed to efficiently identify neighboring nodes within the influence radius, significantly reducing the computational burden during matrix assembly and solution.

The effectiveness of the proposed smoothing methodology in improving boundary quality while preserving geometric features is demonstrated through numerical experiments in Section \ref{sect:num_tests}.
\subsection{Final geometry discretization} \label{sub:Discretization}
\begin{figure}[h!]
    \centering
    \begin{subfigure}[b]{0.21\textwidth}
        \centering
        \includegraphics[height=3.3cm]{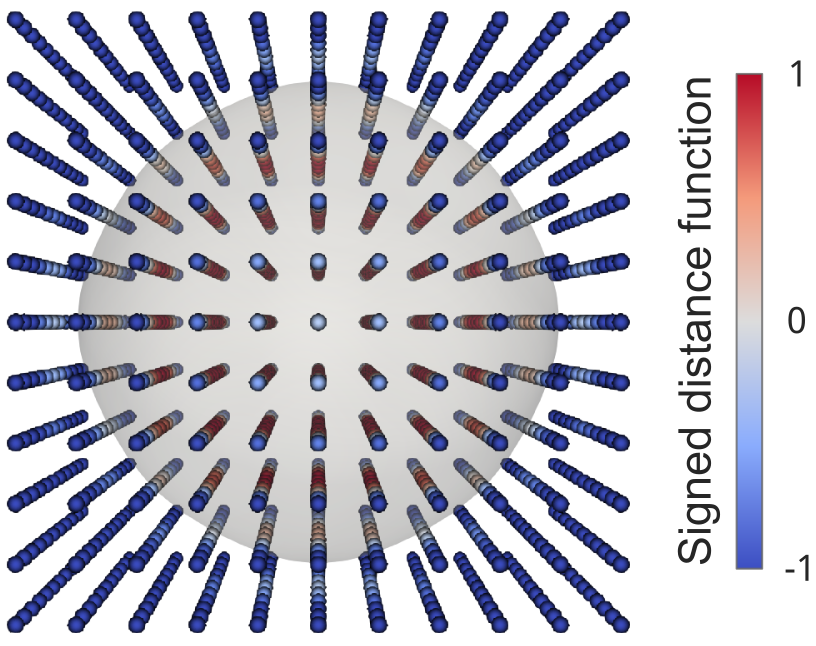}
        \caption{}
        \label{subfig:Sphere_sdf}
    \end{subfigure}
    \hspace{0.02\textwidth} 
    \begin{subfigure}[b]{0.21\textwidth}
        \centering
        \includegraphics[height=3.3cm]{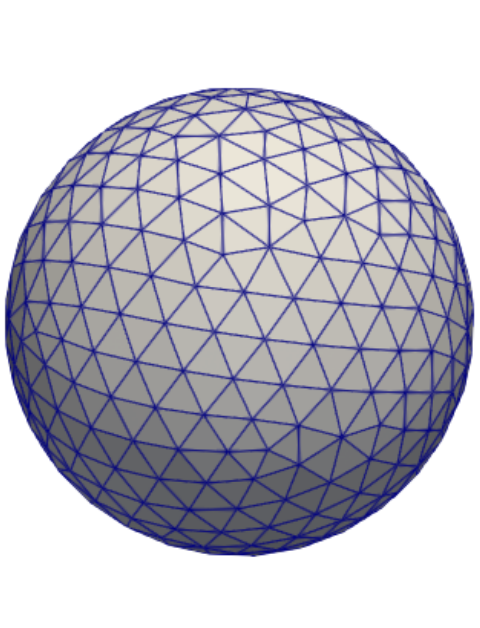}
        \caption{}
        \label{subfig:Sphere_mesh}
    \end{subfigure}
\caption{Geometric representation transition: (a) The continuous SDF representation shown through color-mapped values on a Cartesian grid, with the zero-level contour highlighted in gray, (b) The resulting high-quality tetrahedral mesh generated through isosurface stuffing~\cite{Labelle2007}.}
  \label{fig:Sphere_stuffing}
\end{figure}

\noindent While the signed distance function provides an elegant mathematical representation of the geometry through its zero-level contour (Figure~\ref{subfig:Sphere_sdf}), many engineering applications benefit from a discrete mesh representation for subsequent analysis or manufacturing processes (Figure~\ref{subfig:Sphere_mesh}). Converting the implicit SDF representation to an explicit discrete geometry is therefore an important step in our methodology.

For manufacturing and visualization purposes, surface discretization methods such as Marching Cubes~\cite{Lorensen1987} or Dual Contouring~\cite{Ju2002DualCO} are commonly employed to extract a boundary representation from implicit functions. However, for structural analysis and numerical validation of post-processed geometries, volumetric discretization is essential. While conventional approaches like Delaunay tetrahedralization~\cite{Cheng2003} offer viable solutions, we suggest the Isosurface Stuffing Algorithm~\cite{Labelle2007} for its distinct advantages: guaranteed quality of tetrahedral elements with bounded dihedral angles, adaptive resolution capabilities near boundaries, and ability to handle thin features typical in topology optimization results.
The algorithm efficiently processes the implicit function by directly generating tetrahedral elements without intermediate boundary reconstruction steps, significantly reducing computational complexity making it particularly suitable for large-scale optimization problems.

\section{Numerical tests and validation}\label{sect:num_tests}
This chapter presents a thorough evaluation of the proposed methodology to demonstrate its effectiveness and robustness. We have designed four distinct test cases, each addressing specific aspects of the method's performance and practical applicability.

First, we verify the mathematical correctness of the SDF construction, which forms the foundation of our geometry extraction approach. Second, we investigate the influence of Cartesian grid resolution on result quality, establishing guidelines for practical implementation. The third test evaluates both volume preservation capabilities and the geometric quality of extracted structures through topology optimization objective function evaluation. Finally, we demonstrate the method's practical applicability through a real-world case study of a robot gripper design.

\subsection{Validation of signed distance function generation}
\noindent To validate the accuracy and robustness of the SDF generation algorithm, we designed a simplified yet thorough test case that examines all projection scenarios described in Section \ref{subsub:distance_function_formulation} and detailed in Appendix \ref{subapp:boundary_classes}. The test geometry consists of two finite elements sharing a common face, configured to produce a roof-like structure when using an isocontour value of $\rho_t = 0.5$, as illustrated in Figure~\ref{subfig:strecha_nodal_dense}. The test design was motivated by the difficulties that similar scenarios pose for conventional normal vector-based sign assignment methods.

\subsubsection{Test case configuration}\label{subsub:roof-configuration}
\begin{figure*}[h!]
    \centering
    \begin{subfigure}[b]{0.28\textwidth}
        \centering
        \includegraphics[width=\textwidth]{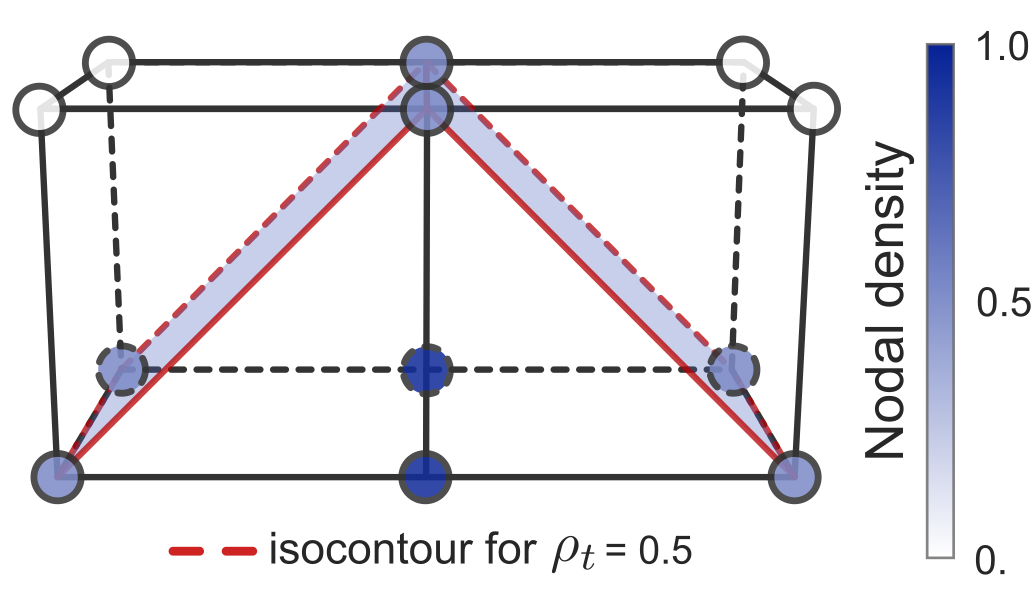}
        \caption{}
        \label{subfig:strecha_nodal_dense}
    \end{subfigure}
    \hspace{0.02\textwidth}
    \begin{subfigure}[b]{0.3\textwidth}
        \centering
        \includegraphics[width=\textwidth]{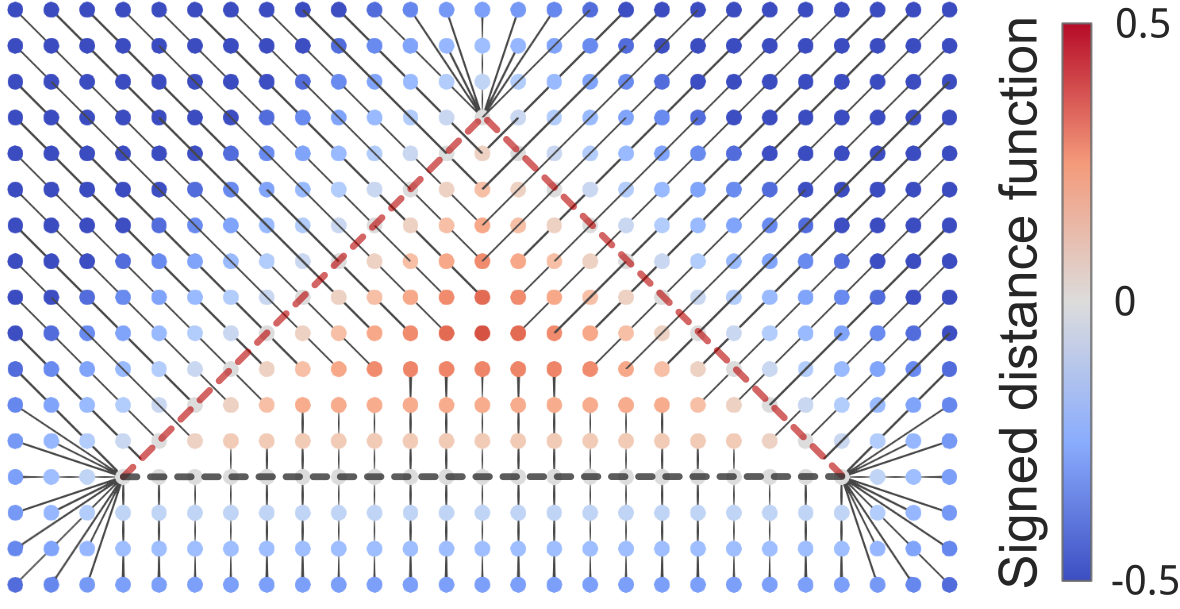}
        \caption{}
        \label{subfig:strecha_sdf}
    \end{subfigure}
    \hspace{0.02\textwidth}
    \begin{subfigure}[b]{0.3\textwidth}
        \centering
        \includegraphics[width=\textwidth]{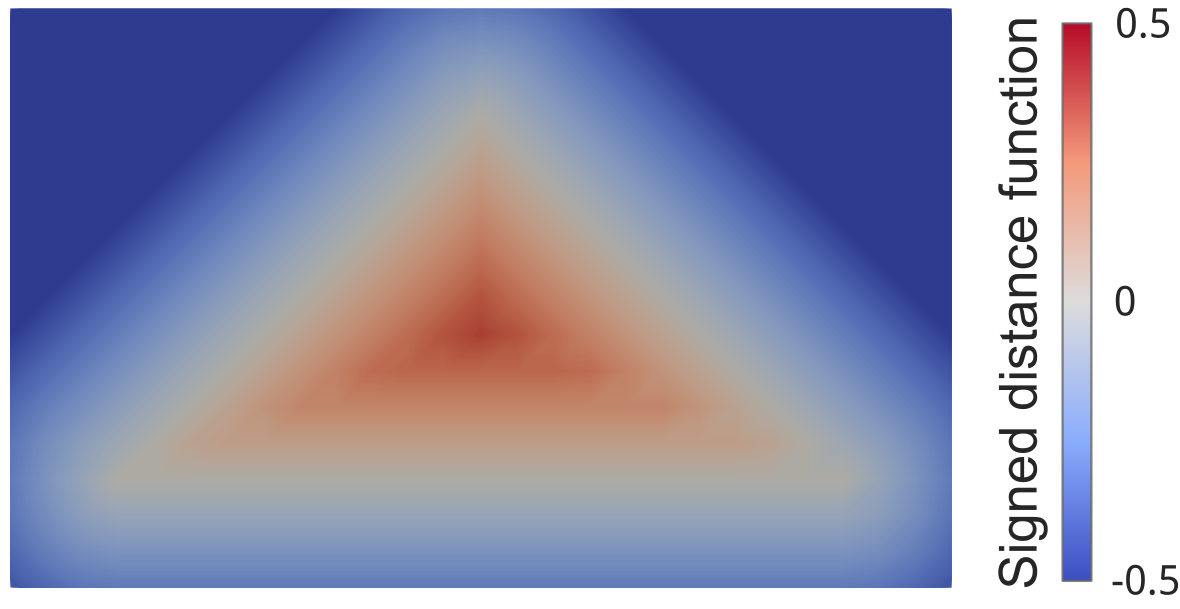}
        \caption{}
        \label{subfig:strecha_sdf_face}
    \end{subfigure}
    \caption{Visual validation of the signed distance function construction. (a) Two finite elements sharing a common face. The nodal densities of these two elements are intentionally assigned to create a roof-like isocontour at $\rho_t = 0.5$, with dark blue representing $\rho_n = 1$, light blue $\rho_n = 0.5$, and white $\rho_n = 0$. A red dashed line indicates the resulting isocontour. (b) Projection analysis showing the computed shortest distances from regular grid nodes to the material domain (defined by the isocontour where interpolated density equals $\rho_t$, or to element faces where interpolated density is higher than $\rho_t$). Gray lines represent projection paths, and the color scale indicates the computed signed distance values. (c) Continuous SDF field obtained by interpolating the discrete nodal distance values within elements, demonstrating smooth geometric transitions and $C^0$ continuity across element boundaries.}
    \label{fig:strecha_nodal4sdf}
\end{figure*}

\noindent Figure~\ref{subfig:strecha_sdf} visualizes the projection of regular grid nodes onto the material domain boundary (isocontour). For clarity of presentation, we display a single layer of grid points projected onto the frontal view of the two-element configuration. The black lines indicate the shortest distance paths computed from each grid point to either the isocontour or element faces where the interpolated density exceeds $\rho_t$.

The analysis demonstrates that our algorithm robustly identifies the closest projection point for each grid node, successfully handling cases with multiple equidistant projections along the symmetry line. Furthermore, Figure~\ref{subfig:strecha_sdf} shows that the sign assignment procedure effectively differentiates between interior and exterior regions of the material domain, ensuring accurate topology representation. This validation is particularly significant given that sharp geometric features typically pose challenges for conventional normal vector-based sign assignment methods~\cite{Baerentzen2005}, yet our density-based approach maintains accuracy even in these geometrically complex regions.

The computed signed distance values demonstrate geometric consistency across the computational domain, as evidenced by the smooth color gradients shown in Figure~\ref{subfig:strecha_sdf_face}. The resulting distance field achieves $C^0$ continuity across element boundaries while accurately capturing geometric features, particularly in regions of high curvature such as the roof ridge. This test case validates the algorithm's capability to correctly detect different boundary types, as elements containing isocontours may simultaneously possess external faces that also require SDF calculation.

\subsection{Grid resolution study}\label{sub:grid_study}
\noindent A spherical test geometry was selected to evaluate the influence of grid resolution on geometric feature extraction due to its smooth, continuous surface characteristics and well-defined analytical solution. To generate the spherical geometry, nodal density values are assigned linearly from $\rho_n = 0$ at mesh corners to $\rho_n = 1$ at the center, then interpolated within elements using finite element shape functions. The sphere geometry emerges as the isocontour where this interpolated density field equals the threshold value $\rho_t = 0.5$.

To assess the discretization independence of our approach, we examine the influence of grid density across four different discretization schemes: uniform hexahedral mesh, non-uniform hexahedral mesh with local refinement, uniform tetrahedral mesh, and non-uniform tetrahedral mesh with local refinement. This testing is desirable? because practical topology optimization applications typically employ irregular meshes and mixed element types, requiring validation across diverse discretization scenarios encountered in engineering practice.

\subsubsection{Resolution analysis}\label{subsub:grid-analysis}
\begin{figure*}[h!]
    \centering
    
    \includegraphics[height=3.5cm]{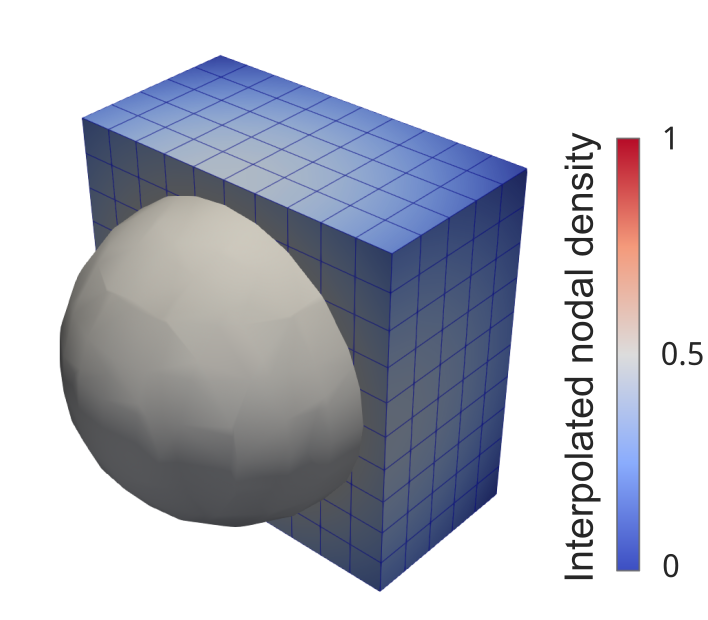}\hspace{0.02\textwidth}
    \includegraphics[height=3.5cm]{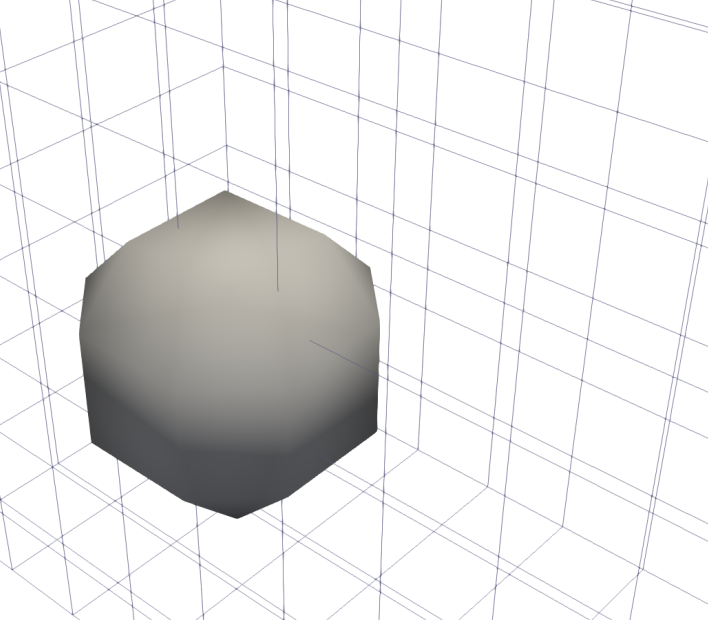}\hspace{0.02\textwidth}
    \includegraphics[height=3.5cm]{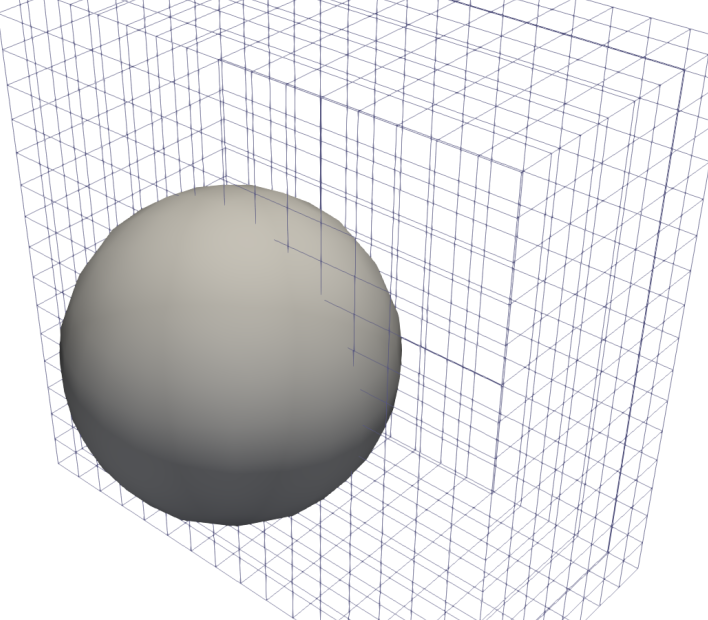}\hspace{0.02\textwidth}
    \includegraphics[height=3.5cm]{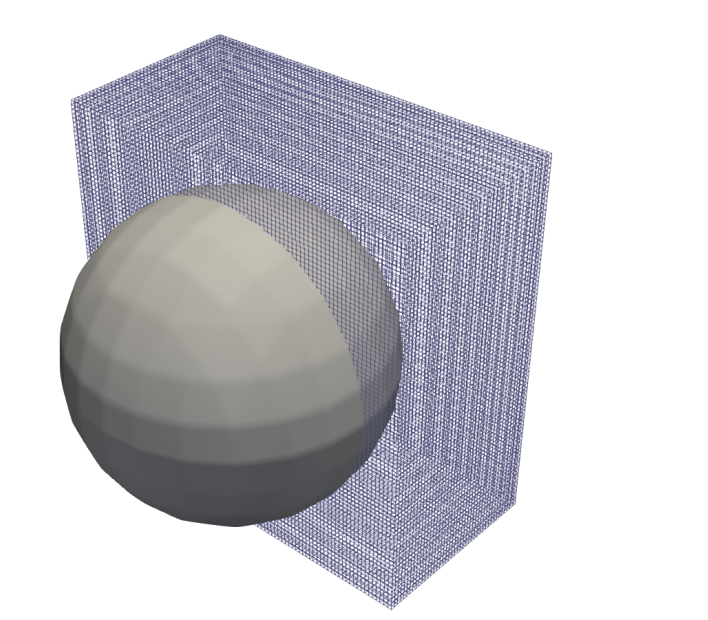}
    
    \vspace{0.2cm}
    \textbf{(a) Uniform hexahedral mesh}
    \vspace{0.5cm}
    
    \includegraphics[height=3.5cm]{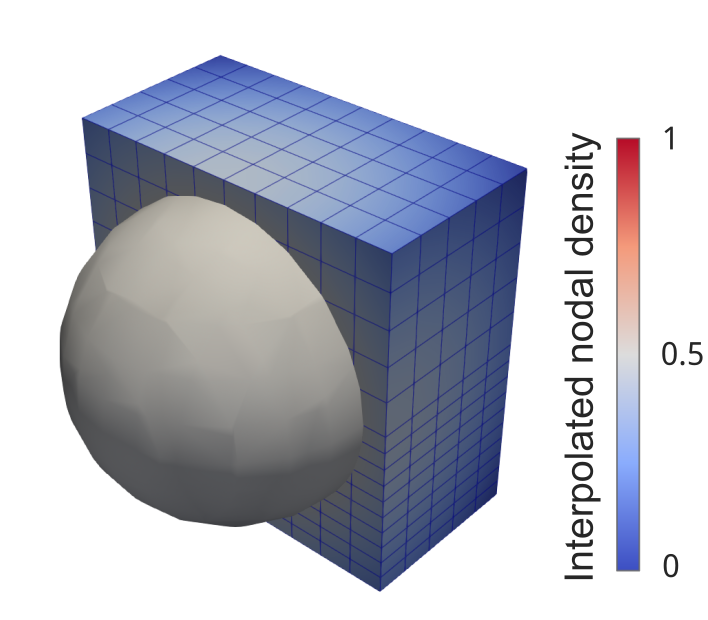}\hspace{0.02\textwidth}
    \includegraphics[height=3.5cm]{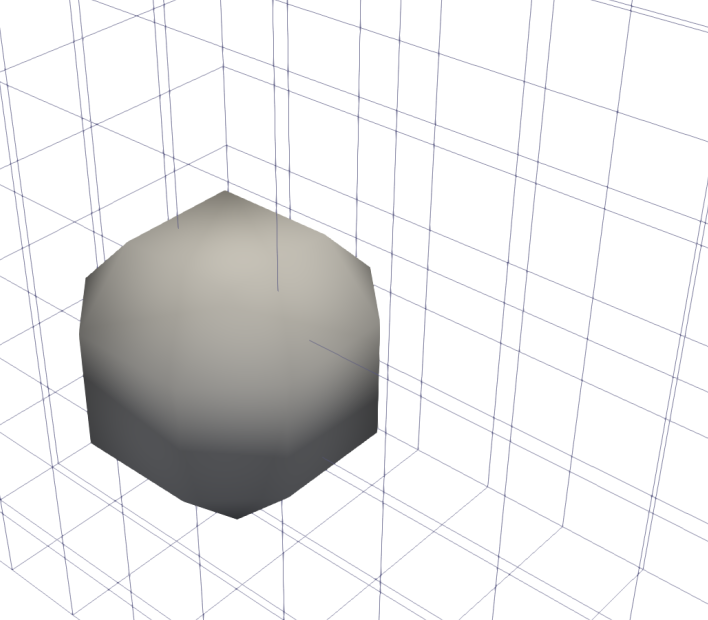}\hspace{0.02\textwidth}
    \includegraphics[height=3.5cm]{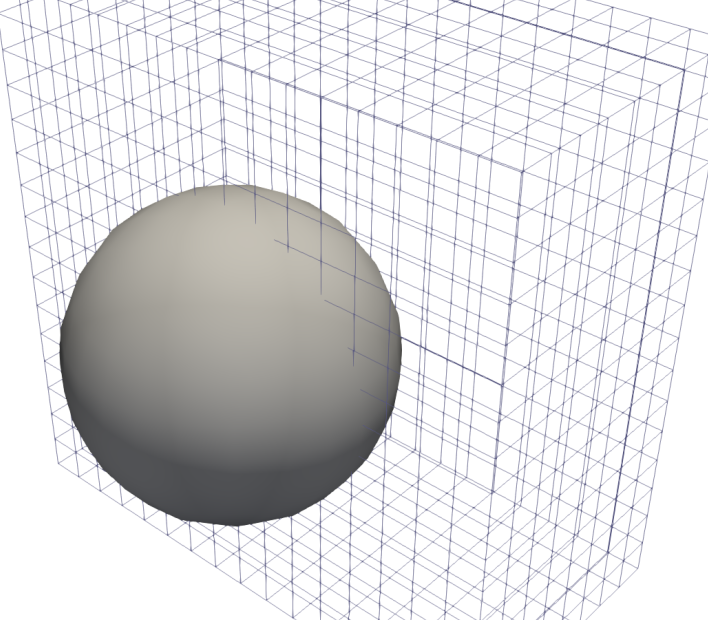}\hspace{0.02\textwidth}
    \includegraphics[height=3.5cm]{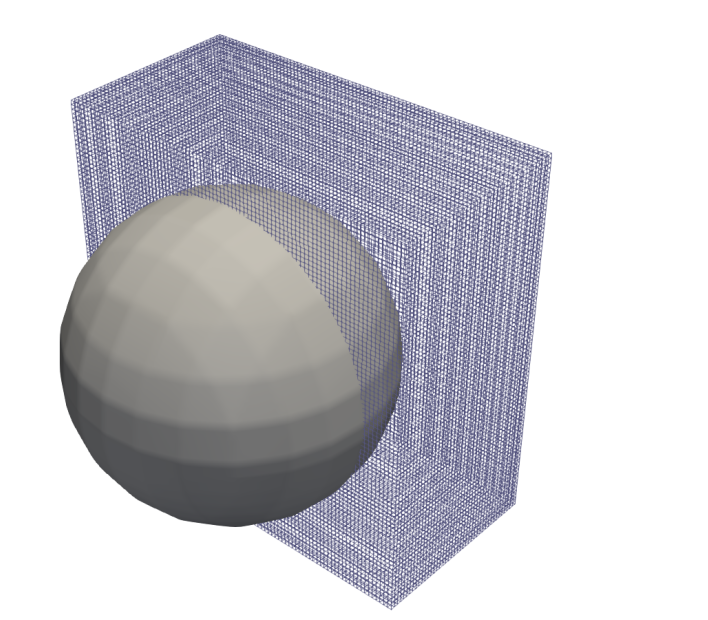}
    
    \vspace{0.2cm}
    \textbf{(b) Non-uniform hexahedral mesh (refined bottom half)}
    \vspace{0.5cm}
    
    \includegraphics[height=3.5cm]{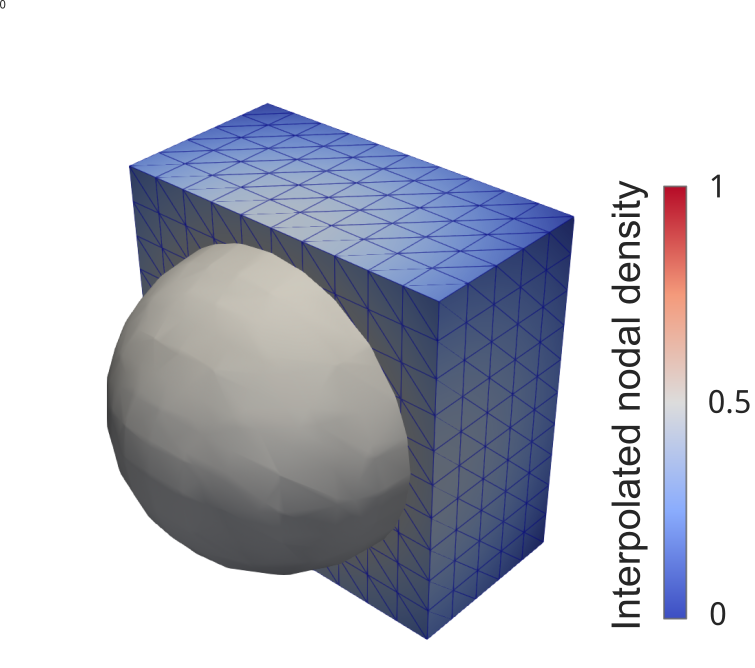}\hspace{0.02\textwidth}
    \includegraphics[height=3.5cm]{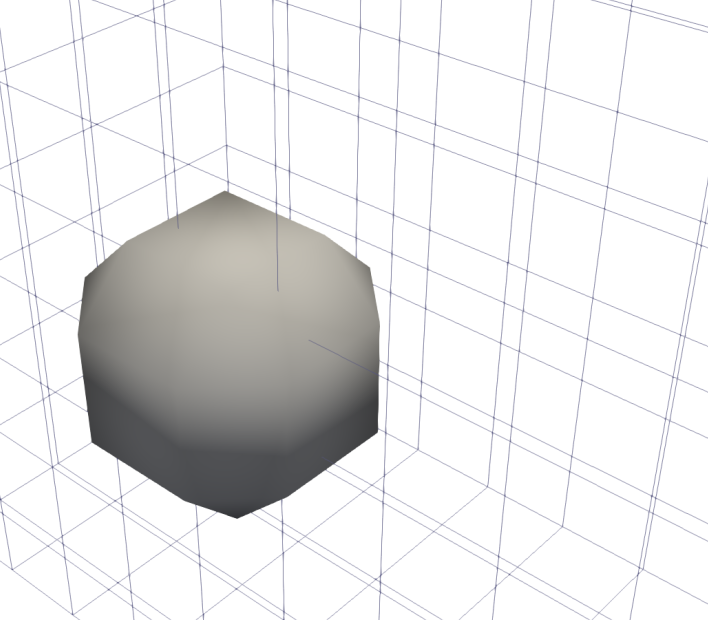}\hspace{0.02\textwidth}
    \includegraphics[height=3.5cm]{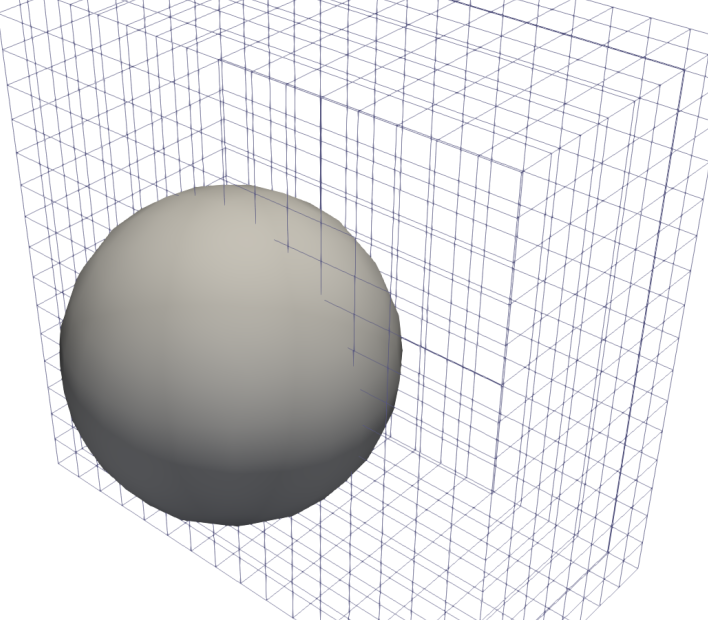}\hspace{0.02\textwidth}
    \includegraphics[height=3.5cm]{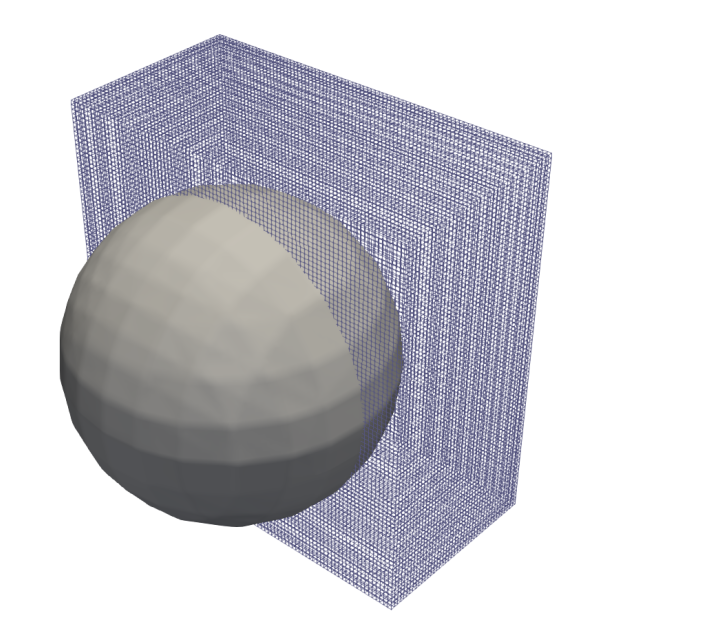}
    
    \vspace{0.2cm}
    \textbf{(c) Uniform tetrahedral mesh}
    \vspace{0.5cm}
    
    \includegraphics[height=3.5cm]{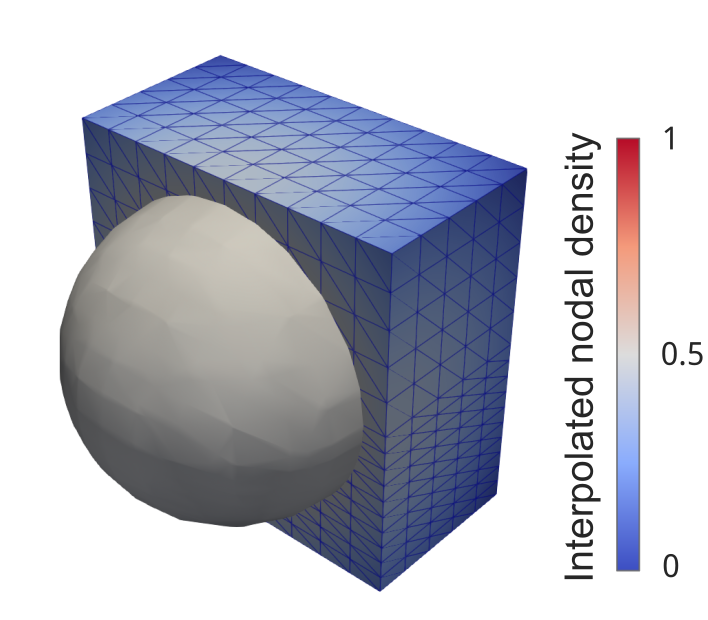}\hspace{0.02\textwidth}
    \includegraphics[height=3.5cm]{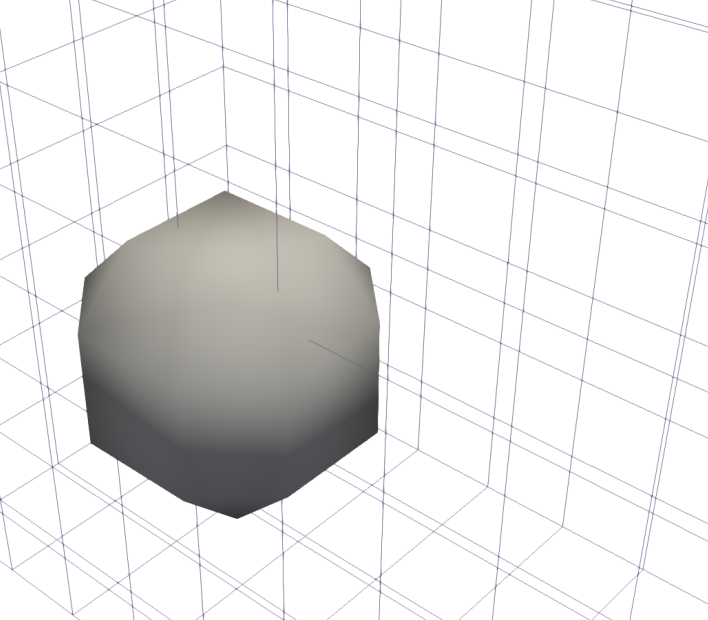}\hspace{0.02\textwidth}
    \includegraphics[height=3.5cm]{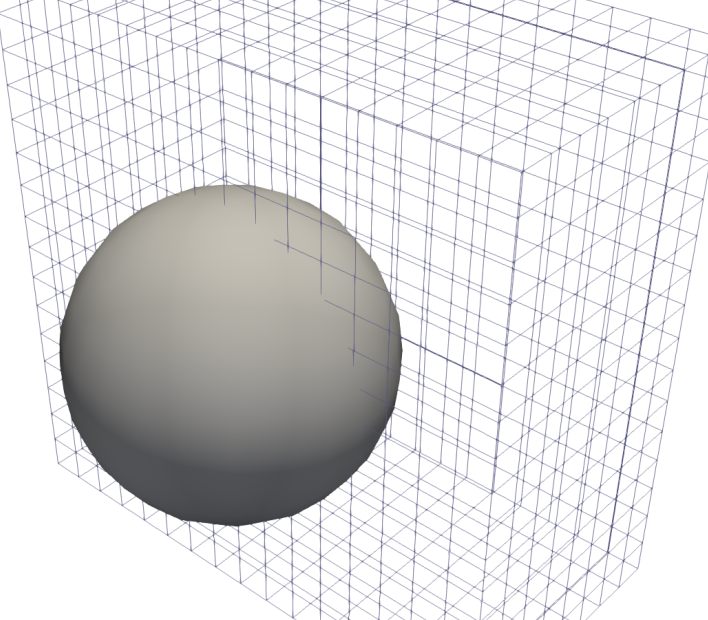}\hspace{0.02\textwidth}
    \includegraphics[height=3.5cm]{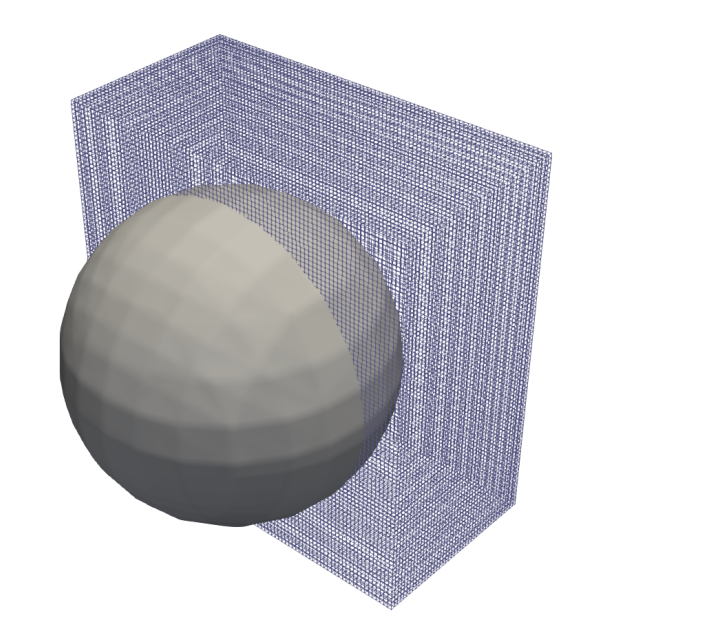}
    
    \vspace{0.2cm}
    \textbf{(d) Non-uniform tetrahedral mesh (refined bottom half)}
    \vspace{0.5cm}

    \caption{Analysis of grid type and resolution effects on geometric feature extraction from different mesh types. Each row shows results for a different initial mesh discretization: (a) uniform hexahedral, (b) non-uniform hexahedral, (c) uniform tetrahedral, and (d) non-uniform tetrahedral meshes. Columns show: initial mesh with nodal densities varying from $\rho_n = 0$ at corners to $\rho_n = 1$ at center with spherical isocontour at $\rho_t = 0.5$, followed by geometric reconstruction using Cartesian grids of different resolutions. All grid spacings are normalized to the uniform hexahedral mesh element size (row a), which serves as the baseline reference. The series demonstrates the critical relationship between grid resolution choice and final geometric accuracy.}
    \label{fig:comprehensive_sphere_analysis}
\end{figure*}

\noindent To evaluate the relationship between Cartesian grid resolution and geometric quality, we conducted a comparative analysis using three distinct grid configurations across all four mesh types, as illustrated in Figure~\ref{fig:comprehensive_sphere_analysis}.\\

\noindent \textbf{Coarse grid configuration} (spacing 3.3 times larger than the baseline hexahedral mesh): Across all mesh types, the coarse grid demonstrates insufficient geometric resolution in the SDF representation. This consistently results in poor geometry representation due to the limited number of grid nodes containing meaningful SDF information, independent of the original mesh element type or local refinement.\\

\noindent \textbf{Reference grid configuration} (spacing matched to the baseline hexahedral mesh): All mesh types achieve optimal geometric results when using this reference grid spacing. The method successfully captures the spherical geometry with appropriate smoothness. Notably, local refinement in the original mesh has no observable effect on the extracted geometry when the grid spacing is appropriately selected.\\

\noindent \textbf{Fine grid configuration} (spacing 8 times smaller than the baseline mesh): The behavior observed with excessively fine discretization is consistent across all mesh types. When the grid spacing becomes significantly smaller than the characteristic element dimensions, the method loses its intended smoothing capability and begins to faithfully preserve individual element isocontours from the original mesh.

\subsubsection{Practical implementation guidelines}\label{subsub:grid-guidelines}
\noindent The results indicate that optimal performance is achieved when the Cartesian grid spacing approximately matches the original finite element mesh size. This finding provides a practical guideline for implementing the method: the Cartesian grid resolution should be selected based on the characteristic element size of the original topology optimization mesh. For uniform meshes, the grid spacing should match the typical element edge length, while for locally refined meshes, the grid spacing should be based on the predominant element size in regions containing significant structural features. This approach ensures proper geometric information transfer while maintaining appropriate levels of smoothing and feature preservation.

\subsubsection{Grid convergence error analysis}\label{subsub:grid-convergence}
\noindent To validate the numerical accuracy of our SDF-based geometry extraction method, we examine its convergence characteristics through mesh refinement study.

For this convergence study, we select a spherical test geometry that provides an ideal validation case due to its smooth analytical boundary and known exact volume. We define a cube with edge length 2, centered at the origin, discretized using uniform hexahedral meshes of varying resolution. Each mesh node is assigned a signed distance value corresponding to its shortest distance from a sphere centered at the origin with radius 0.5. The convergence metric evaluates the relative volume error between the analytical sphere volume ($V_{\text{analytical}} = \frac{4}{3}\pi r^3$) and the computed volume from the discrete isocontour representation interpolated using linear finite element shape functions.

\begin{figure}[h!]
    \centering
    \includegraphics[width=0.45\textwidth]{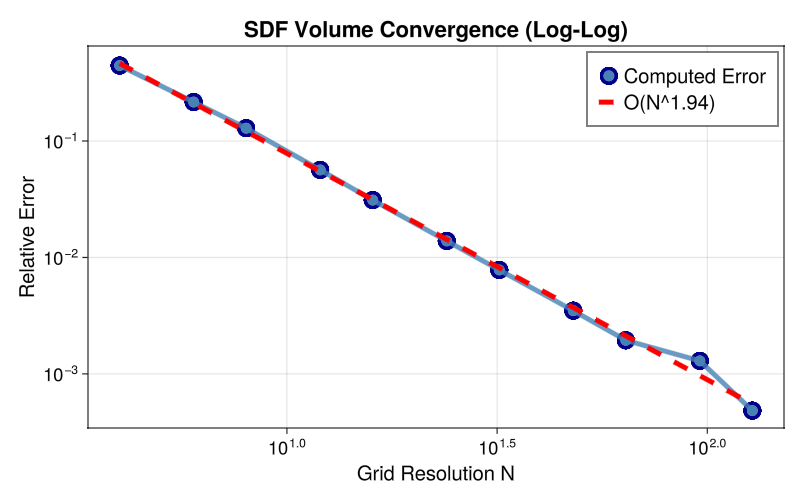}
    \caption{Convergence analysis demonstrating relative volume error reduction with mesh refinement for spherical test geometry, confirming second-order accuracy ($p \approx 1.94$).}
    \label{fig:convergence_analysis}
\end{figure}

We analyze six discretization levels ranging from $N = 4$ to $N = 128$ elements per edge direction. The analysis of the method convergence rate is shown in Figure~\ref{fig:convergence_analysis}. The obtained spatial order of $p \approx 1.94$ confirms the near-optimal second-order behavior.

\subsection{Volume preservation and geometric quality assessment}\label{sub:cantilever}
\subsubsection{Test case definition}\label{subsub:cantilever-definition}
\noindent To evaluate the geometric quality of results obtained through the proposed post-processing method and validate its volume preservation capabilities, we examine a cantilever beam test case with point loading at its free end - one of the classical benchmark problems in topology optimization literature~\cite{Tovar2014}. The beam geometry, adopted from~\cite{Sotola2024}, provides an ideal validation case that combines several advantageous characteristics: its quasi-2D behavior enables clear result interpretation, it produces diverse structural features including both thick and thin members with various interconnections, and it maintains simplicity in problem formulation.

\begin{figure}[h!]
    \centering
    \includegraphics[width=0.3\textwidth]{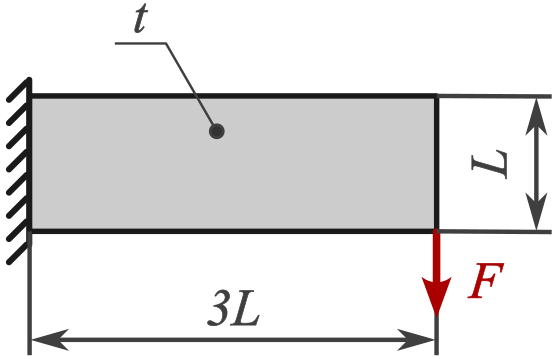}
    \caption{Cantilever beam test case: geometry and boundary conditions.}
    \label{fig:beam_def}
\end{figure}

The beam has a characteristic length $L = 20$ mm and thickness $t = 4$ mm. The boundary conditions, illustrated in Figure~\ref{fig:beam_def}, specify that the beam is fixed along its left edge and subjected to a vertical force of $1$\,N at its bottom-right corner. For the optimization process, the volume fraction constraint is set to $40\%$ of the initial design space. The material properties are defined by Young's modulus $E = 1$ MPa and Poisson's ratio $\nu = 0.3$. The domain is discretized using a hexahedral mesh with uniform element size of 1 mm. The topology optimization employs filtering of design variables (density) and incorporates SIMP penalization with $p = 3$ to suppress intermediate densities, see~\eqref{eqn:mod_simp}.

\subsubsection{Linear post-processing approach}\label{subsub:cantilever-linear}
\noindent This approach uses the nodal density values obtained through the mapping procedure described in Section \ref{sect:methodology} (Step 2) and applies Paraview's surface extraction module with the Contour filter at an iso-value of 0.5186. This threshold value was iteratively determined to maintain the 40\% volume fraction from the original topology optimization results. The Paraview module generates a triangulated surface mesh by identifying element interfaces where the density field crosses the threshold value and determining vertex positions through linear interpolation. To enable numerical analysis, a volumetric mesh was created from this surface representation using Tetgen~\cite{Si2015}, which generates tetrahedral volume meshes using Delaunay triangulation algorithms.

\subsubsection{Results assessment and validation}\label{subsub:cantilever-results}
\begin{figure*}[h!]
    \centering
    \begin{subfigure}[b]{0.45\textwidth}
        \centering
        \includegraphics[width=\textwidth]{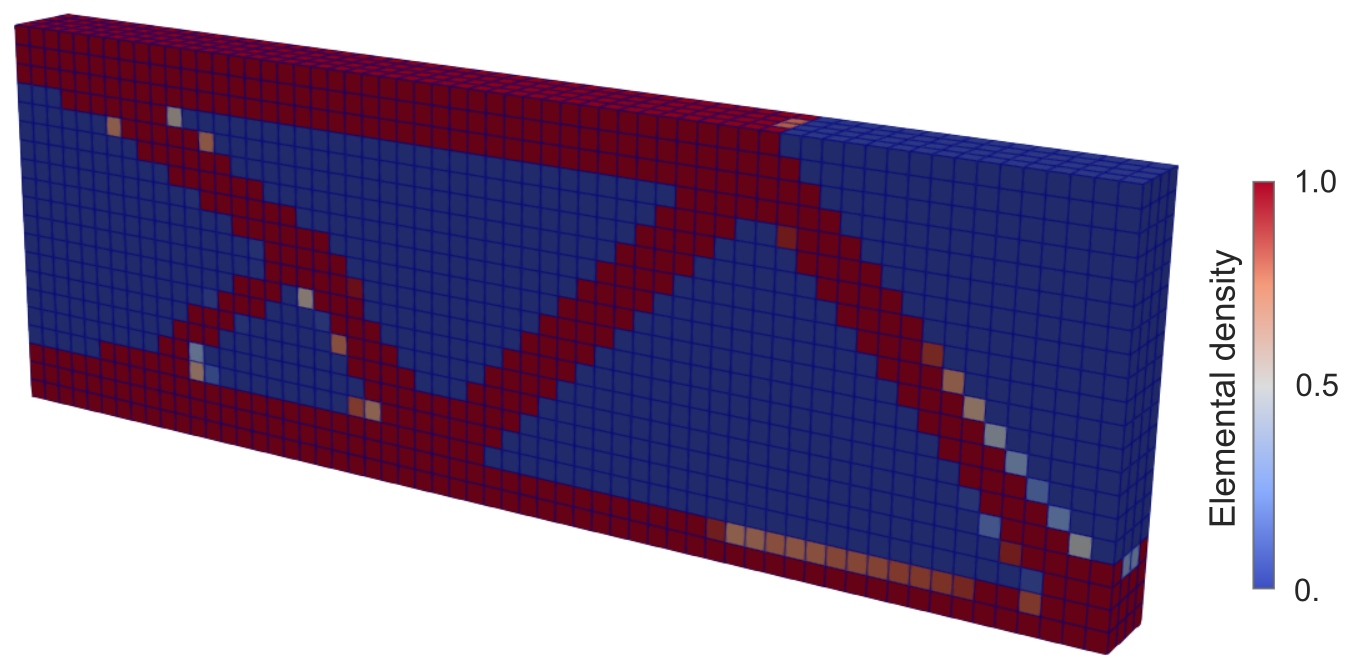}
        \caption{}
        \label{subfig:beam_SIMP}
    \end{subfigure}
    \hspace{0.05\textwidth}
    \begin{subfigure}[b]{0.41\textwidth}
        \centering
        \includegraphics[width=\textwidth]{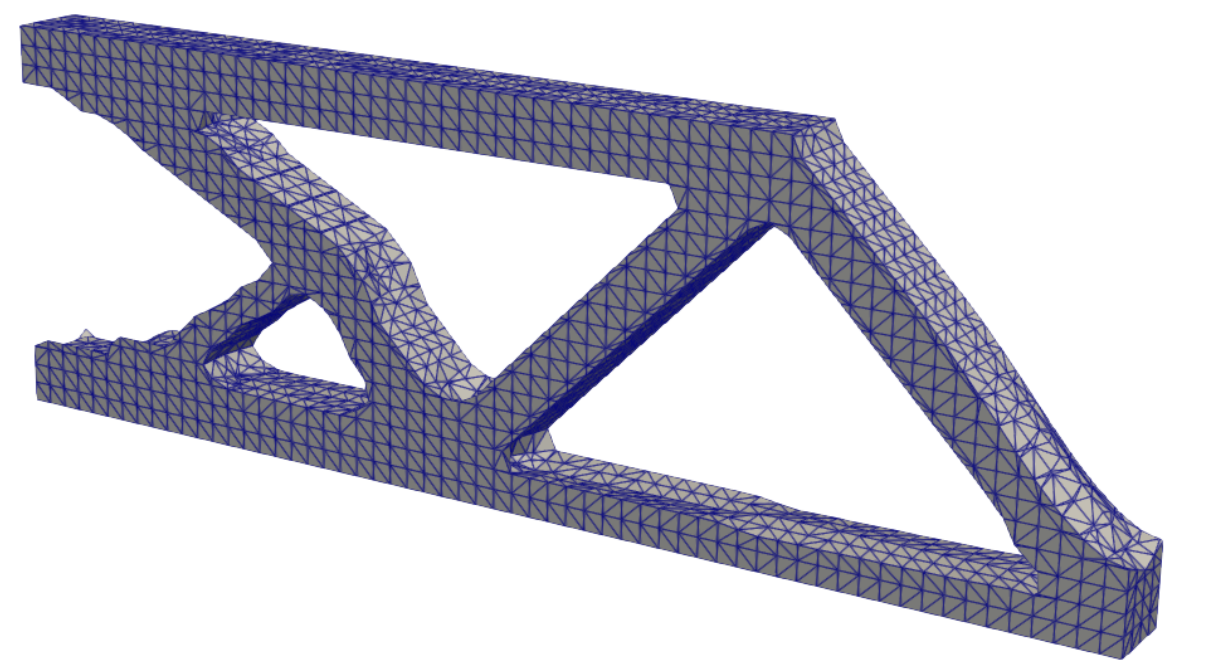}
        \caption{}
        \label{subfig:beam_linear}
    \end{subfigure}
    \hspace{0.05\textwidth}
    \begin{subfigure}[b]{0.41\textwidth}
        \centering
        \includegraphics[width=\textwidth]{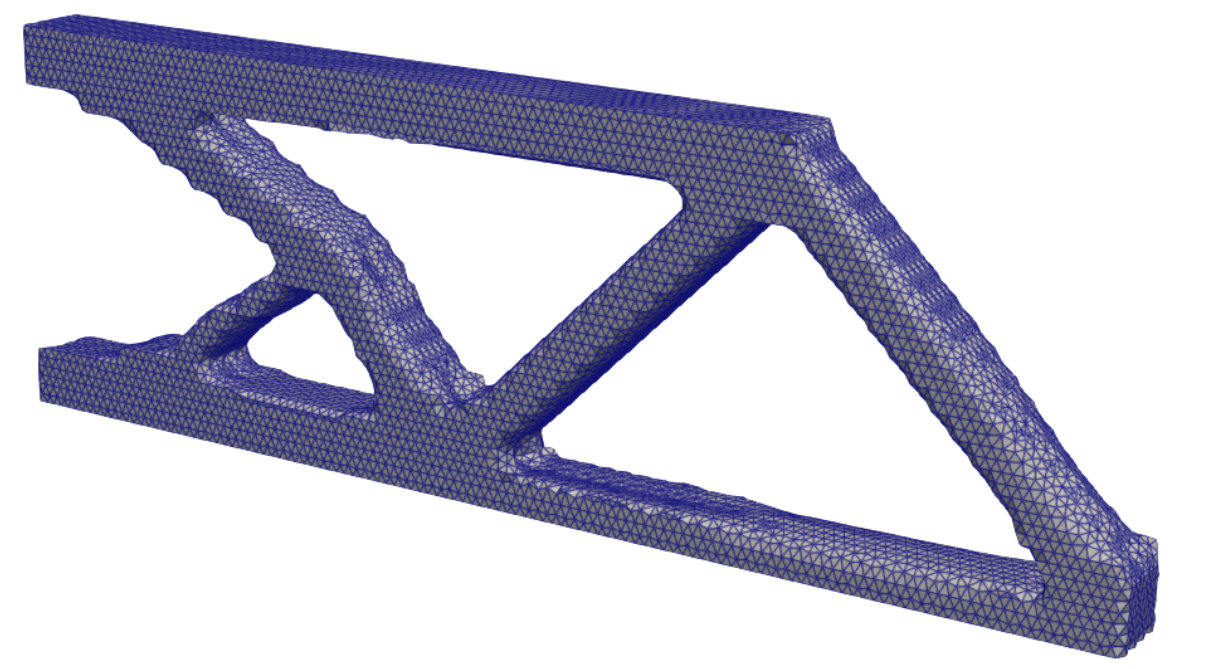}
        \caption{}
        \label{subfig:beam_sdf}
    \end{subfigure}
    \caption{Comparative visualization of the optimized cantilever beam structure across different processing stages: (a) raw SIMP topology optimization result exhibiting characteristic intermediate densities and jagged boundaries, (b) the result of conventional post-processing, which yields a piecewise linear boundary representation, and (c) proposed SDF-based method result demonstrating smooth boundaries while preserving key geometric features and maintaining volume constraints.}
    \label{fig:beam_pp}
\end{figure*}

\noindent To evaluate the proposed method's capabilities, we analyze and compare three distinct versions of the optimized structure: the original SIMP result (Figure~\ref{subfig:beam_SIMP}), results from conventional linear post-processing (Figure~\ref{subfig:beam_linear}), and outcomes from our proposed SDF-based method (Figure~\ref{subfig:beam_sdf}).

The validation metrics were selected to evaluate four critical aspects:
\begin{enumerate}
\item Volumetric accuracy through volume fraction preservation
\item Structural efficiency via strain energy comparison
\item Boundary smoothness through maximum equivalent stress analysis
\item Structural behavior consistency via deformation energy density evaluation
\end{enumerate}

\begin{table*}[h!]
\centering
\renewcommand{\arraystretch}{1.5}
\begin{tabular}{|>{\bfseries}l|c|c|c|c|}
\hline
\rowcolor{gray!20}
\textbf{} & \textbf{Raw SIMP results} & \textbf{Linear post-processing} & \textbf{SDF post-processing}\\
\hline
\textbf{Volume fraction} [\si{1}] & 0.4007 & 0.4017 & 0.3942 \\
\hline
\textbf{Deform. energy} [\si{\joule}]  & 3.411 & 2.832 & 3.075 \\
\hline
\textbf{Max. Eq. stress} [\si{\mega\pascal}] & - & 0.3010 & 0.2543 \\
\hline
\textbf{Deform. energy density} [\si{\mega\joule\per\cubic\metre}] & 1.773 & 1.479 & 1.625 \\
\hline
\end{tabular}
\caption{Comparison of performance metrics for the cantilever beam case study}
\label{tab:comparison}
\end{table*} 

\noindent The volume fraction analysis (see Table \ref{tab:comparison}) indicates excellent constraint satisfaction across all approaches. The SDF post-processing method achieves a volume fraction of 0.3942, which closely aligns with the original SIMP result (0.4007). The slight reduction in volume fraction (approximately 1.6\%) is attributed to discretization effects during the tetrahedral mesh generation, where the discrete elements cannot perfectly capture the implicit geometry boundaries. The conventional linear post-processing method maintains a volume fraction of 0.4017, as expected, since its threshold was specifically calibrated to preserve the target volume ratio.

From a structural performance perspective, the deformation energy analysis reveals nuanced efficiency differences between methods. The linear post-processing approach achieves the lowest deformation energy (2.832 J), indicating superior stiffness compared to both the original SIMP solution (3.411 J) and SDF post-processed geometry (3.075 J). This result is expected as the linear method directly preserves the raw optimization features without additional smoothing operations. The SDF-based method demonstrates a favorable balance between performance and geometry quality—achieving approximately 9.9\% lower deformation energy than the original SIMP solution while simultaneously delivering significantly smoother boundaries. This represents an effective compromise that enhances manufacturability without substantially compromising the structural efficiency inherent in the optimized design.

Most significantly, the maximum equivalent stress analysis demonstrates a clear advantage of the SDF-based approach. The SDF post-processing method shows reduced maximum stress values (0.2543 MPa) compared to the linear post-processing method (0.3010 MPa). The SDF method achieves a 15.5\% reduction in peak stress compared to conventional post-processing. This improvement can be attributed to the smooth boundary transitions generated by the SDF approach, which substantially mitigates artificial stress concentrations typically associated with sharp geometric features in conventional post-processing.

The deformation energy density values further validate these findings, showing that the SDF-based method (1.625e6 J/m³) more closely matches the original SIMP results (1.773e6 J/m³) than the linear post-processing approach (1.479e6 J/m³). This consistency in energy density indicates that the proposed SDF method preserves not only the geometric characteristics but also the mechanical behavior of the optimized structure.

\subsection{Application to complex engineering design: Robot gripper case study}\label{sub:Robot_gripper}
\noindent While simplified benchmark problems provide valuable insights into method performance, real engineering applications often present additional challenges that must be addressed. To validate our approach on a practical industrial case, we analyze the topology optimization and subsequent geometry extraction for a robot gripper design.

\subsubsection{Test case definition}\label{subsub:chapadlo-definition}
\begin{figure}[h!]
    \centering
    \includegraphics[width=0.35\textwidth]{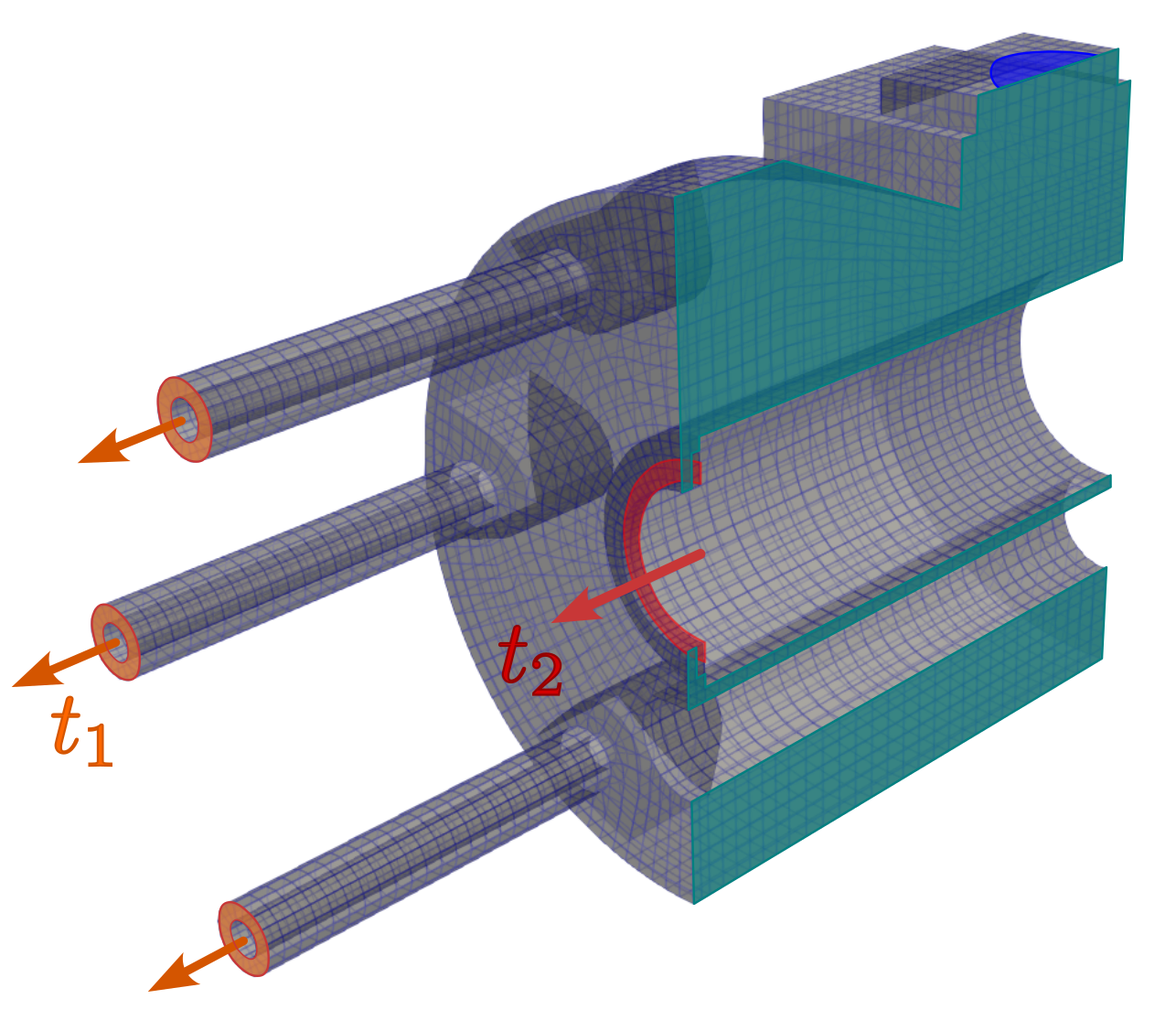}
    \caption{Robot gripper test case: geometry and boundary conditions.}
    \label{fig:chapadlo_def}
\end{figure}

\noindent The optimization problem involves designing a gripper structure subject to combined loading conditions and geometric constraints, as illustrated in Figure~\ref{fig:chapadlo_def}. The design domain consists of a mounting plate with attachment points for suction cups and a camera system. The problem formulation incorporates a single loading scenario that combines three simultaneous load components: distributed pressure $t_1 = 9850$~Pa from lifting a 3~kg workpiece across four suction cups, surface traction $t_2 = 1852$~Pa due to camera and cable weight (400~g), and body forces resulting from vertical acceleration at 6~m/s². Fixed boundary conditions (shown in blue) are applied at the mounting interface, while symmetry conditions along the central plane (shown in turquoise) reduce the computational domain to half of the gripper. The structure utilizes ABS M30 plastic with material properties defined by Young's modulus of 2400~MPa, Poisson's ratio of 0.35, density of 1040~kg/m³, and yield strength of 30~MPa, with the optimization constrained by a volume fraction limit of 0.3.
\subsubsection{Linear post-processing approach}\label{subsub:chapadlo-linear}
\noindent To maintain consistency with the cantilever beam validation methodology, we employed the same conventional linear post-processing approach for this case study. The procedure is identical to the geometry extraction process outlined in Section~\ref{subsub:cantilever-linear}, with the only difference being the iso-value selection of 0.5007 to preserve the target 30\% volume fraction from the original topology optimization results.

\subsubsection{Results assessment and validation}\label{subsub:chapadlo-results}
\begin{figure*}[h!]
    \centering
    \begin{subfigure}[b]{0.28\textwidth}
        \centering
        \includegraphics[width=\textwidth]{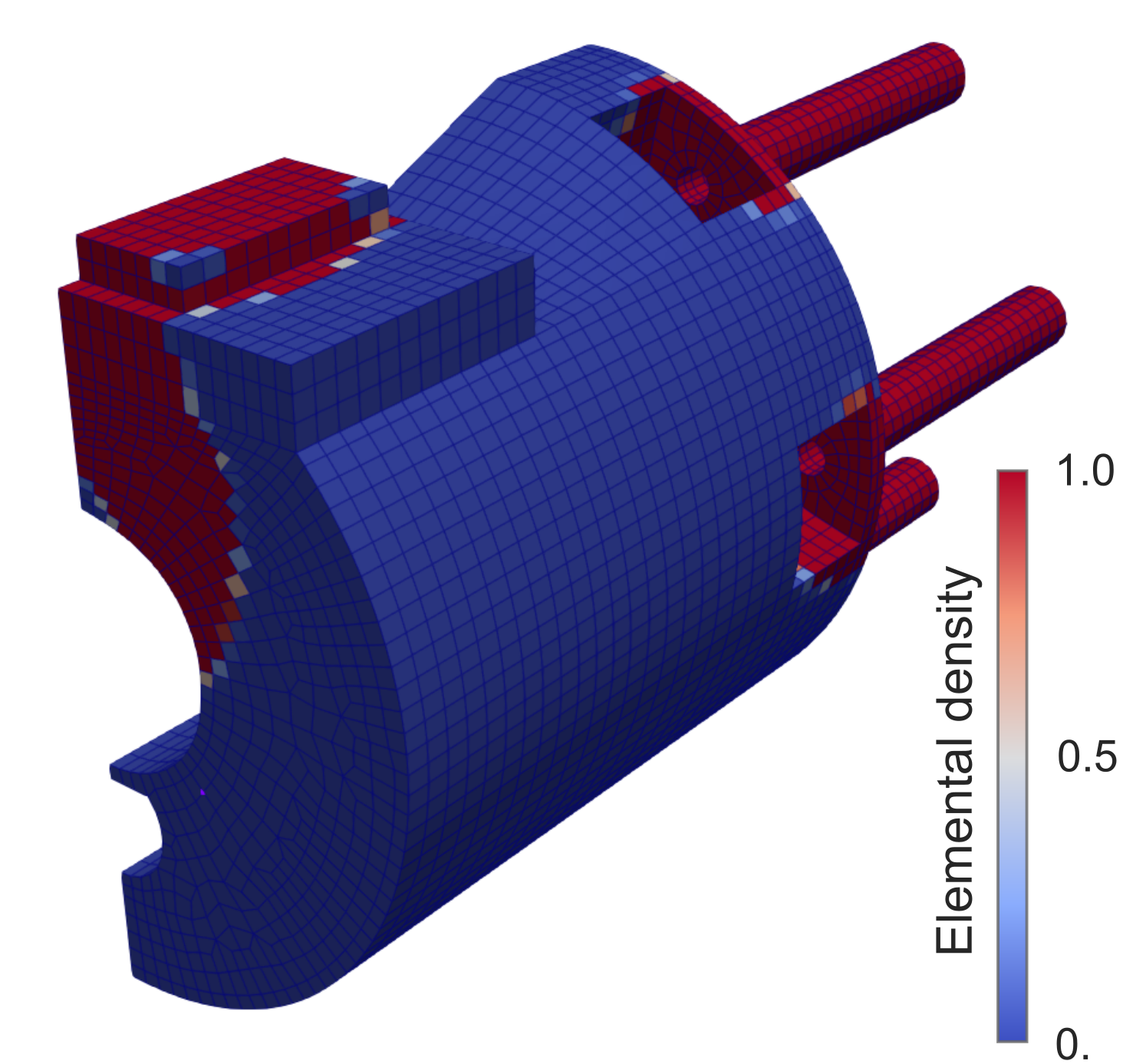}
        \caption{}
        \label{subfig:chapadlo_dense}
    \end{subfigure}
    \hspace{0.05\textwidth}
    \begin{subfigure}[b]{0.28\textwidth}
        \centering
        \includegraphics[width=\textwidth]{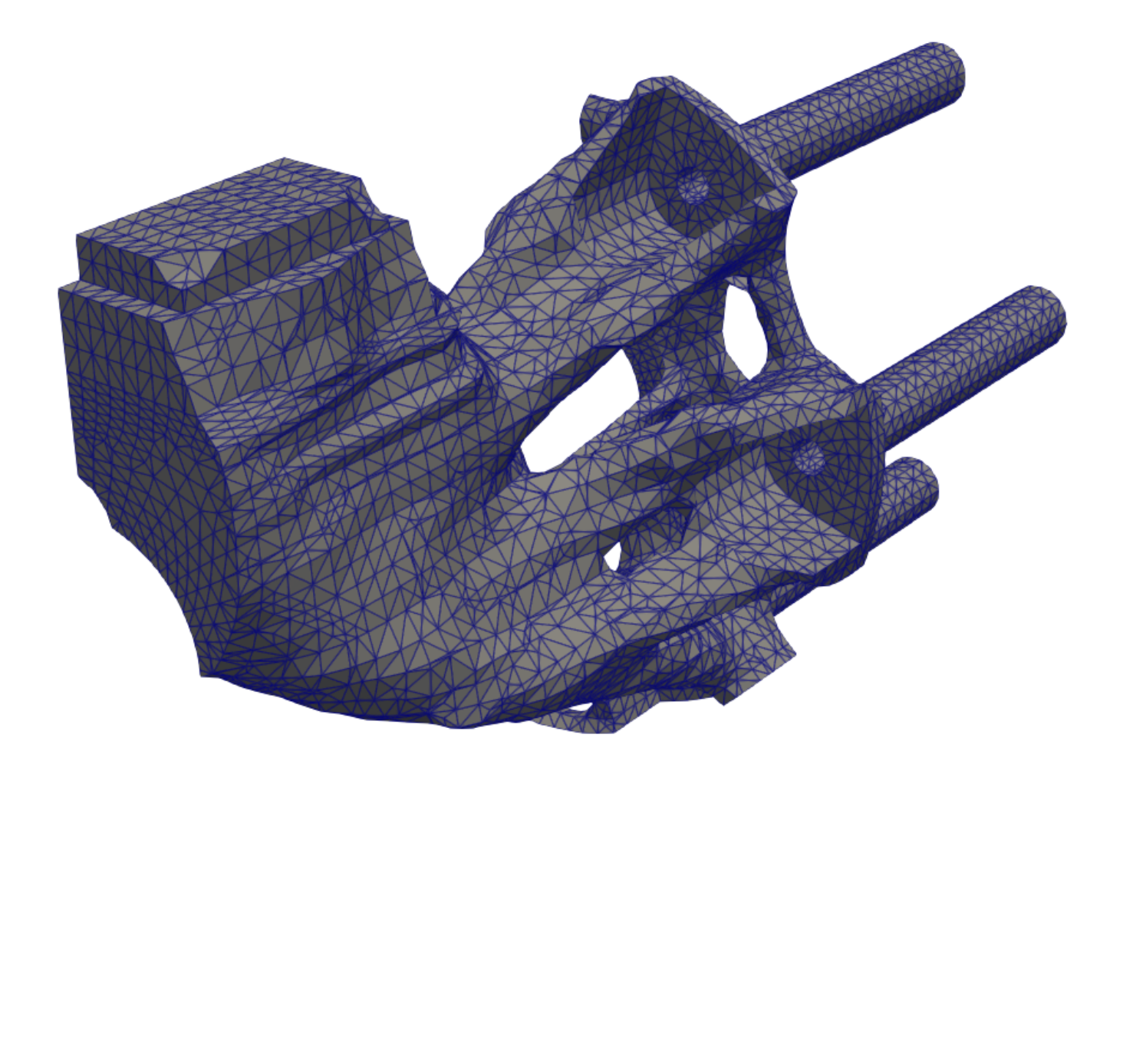}
        \caption{}
        \label{subfig:chapadlo_linear}
    \end{subfigure}
    \hspace{0.05\textwidth}
    \begin{subfigure}[b]{0.28\textwidth}
        \centering
        \includegraphics[width=\textwidth]{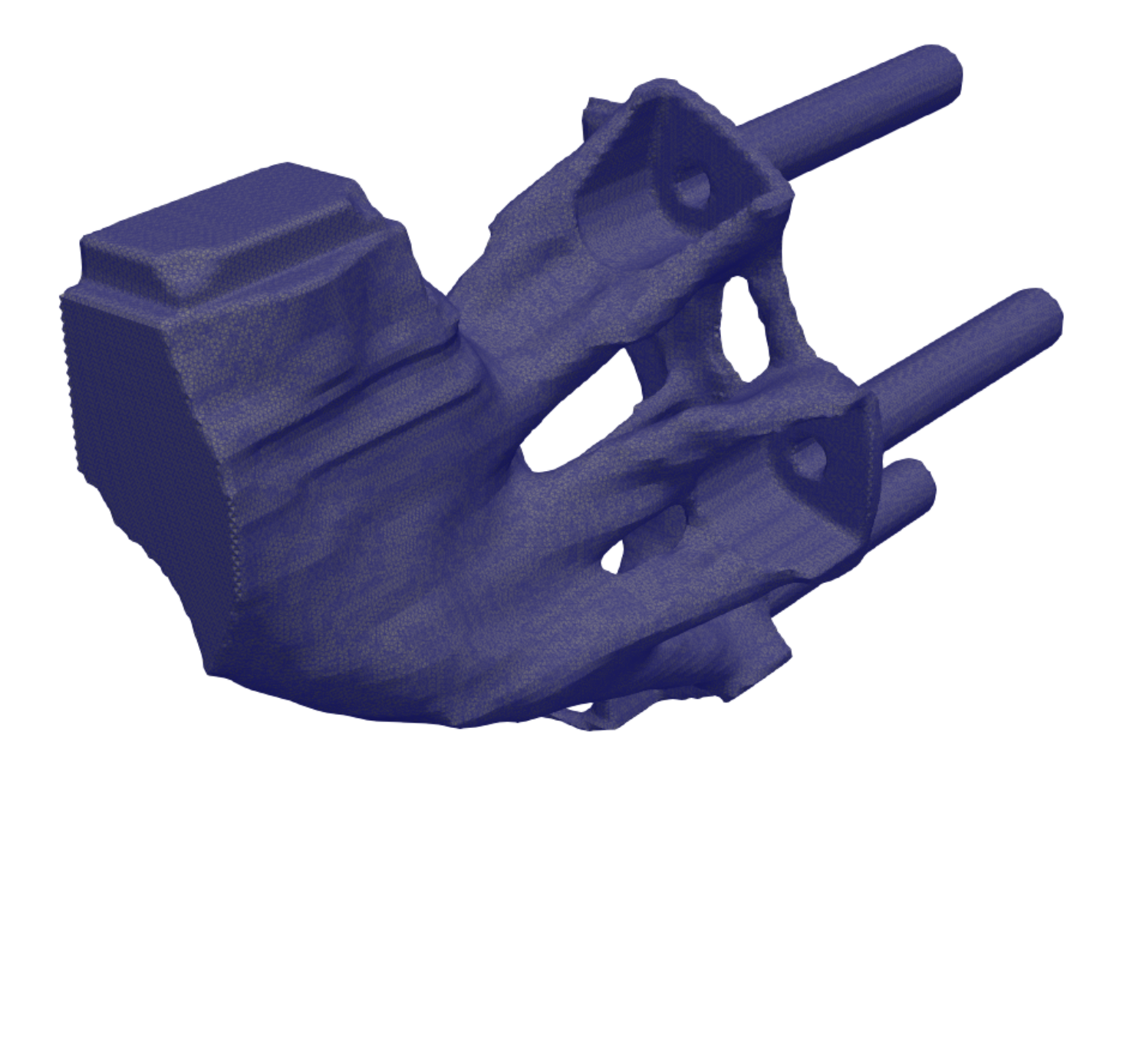}
        \caption{}
        \label{subfig:chapadlo_sdf}
    \end{subfigure}
    \caption{Post-processing results for the robot gripper: (a) Raw element density distribution from topology optimization, (b) Extracted geometry using conventional linear post-processing, (c) Result from proposed SDF-based post-processing method.}
    \label{fig:chapadlo_results}
\end{figure*}

\noindent To evaluate the proposed method's capabilities in a complex engineering context, we analyze and compare three distinct versions of the optimized gripper structure. The original SIMP results (Figure~\ref{subfig:chapadlo_dense}) were obtained using an unstructured mesh comprising 16,490 linear hexahedral elements and 20,542 nodes, displaying the characteristic element-wise constant density distribution with intermediate density regions and stair-stepped interfaces typical of SIMP-based optimization. These results are compared against conventional linear post-processing outcomes (Figure~\ref{subfig:chapadlo_linear}) and our proposed SDF-based method results (Figure~\ref{subfig:chapadlo_sdf}).

The quantitative performance comparison presented in Table~\ref{tab:chapadlo_comparison} demonstrates the effectiveness of the proposed SDF-based post-processing approach for this complex engineering application. Volume fraction preservation is achieved with exceptional accuracy across all methods, with both post-processing approaches maintaining the target volume fraction of 0.3000 within 0.03\% deviation. This precise volumetric control is expected and stems directly from the mathematical construction of the SDF methodology.

The deformation energy analysis reveals structural performance characteristics consistent with the cantilever beam validation. The linear post-processing method achieves the lowest deformation energy (94.8 kJ), reflecting its strict preservation of geometric features that contribute to structural stiffness. The SDF post-processing method exhibits a moderate increase in deformation energy (102.1 kJ), representing approximately 7.7\% higher energy compared to linear post-processing, yet maintaining an 11.6\% improvement over the raw SIMP results (115.6 kJ). This performance profile demonstrates the method's ability to balance structural efficiency with geometric quality enhancement. These findings are confirmed by the deformation energy density values presented in Table~\ref{tab:chapadlo_comparison}, which show that the SDF-based method maintains closer correspondence to the original SIMP distribution compared to linear post-processing.

\begin{table*}[h!]
\centering
\renewcommand{\arraystretch}{1.5}
\begin{tabular}{|>{\bfseries}l|c|c|c|c|}
\hline
\rowcolor{gray!20}
\textbf{} & \textbf{Raw SIMP results} & \textbf{Linear post-processing} & \textbf{SDF post-processing}\\
\hline
\textbf{Volume fraction} [\si{1}] & 0.3000 & 0.3001 & 0.3001 \\
\hline
\textbf{Deform. energy} [\si{\kilo\joule}]  & 115.6 & 94.8 & 102.1 \\
\hline
\textbf{Deform. energy density} [\si{\mega\joule\per\cubic\metre}] & 363.8 & 298.3 & 321.1 \\
\hline
\end{tabular}
\caption{Comparison of performance metrics for the robot gripper case study}
\label{tab:chapadlo_comparison}
\end{table*}
\begin{figure}[h!]
    \centering
    \begin{subfigure}[b]{0.12\textwidth}
        \centering
        \includegraphics[height=2.3cm]{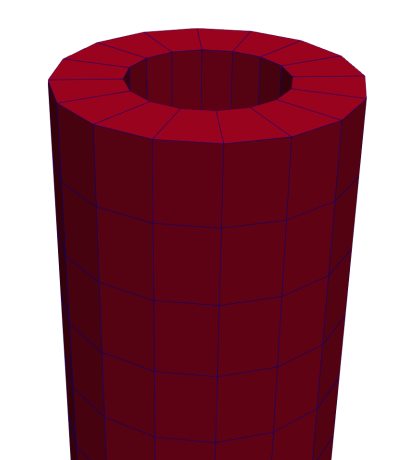}
        \caption{}
        \label{subfig:Leg_raw}
    \end{subfigure}
    \hspace{0.01\textwidth}
    \begin{subfigure}[b]{0.12\textwidth}
        \centering
        \includegraphics[height=2.3cm]{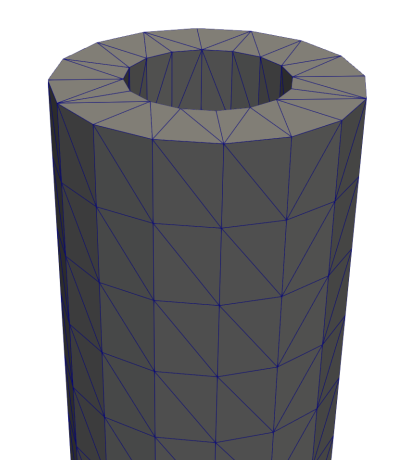}
        \caption{}
        \label{subfig:Leg_linear}
    \end{subfigure}
    \hspace{0.01\textwidth}
    \begin{subfigure}[b]{0.12\textwidth}
        \centering
        \includegraphics[height=2.3cm]{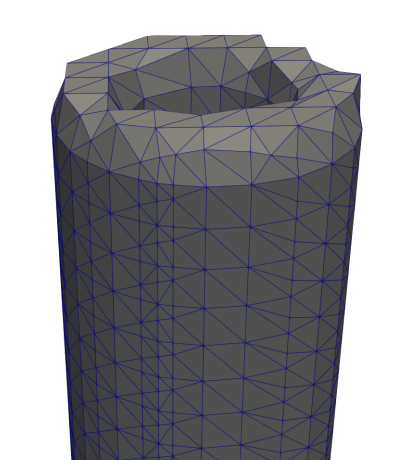}
        \caption{}
        \label{subfig:Leg_sdf}
    \end{subfigure}
\caption{View of one of the critical surfaces for boundary condition application - the gripper leg end. (a) Raw topology optimization results, (b) conventional linear post-processing with piecewise linear boundary representation, (c) proposed SDF-based post-processing showing global smoothing effects that may require careful consideration at boundary condition interfaces, though this can be partially addressed through subsequent discretization strategies.}
  \label{fig:Leg_comparison}
\end{figure}
From a geometric perspective, the comparison of post-processing approaches reveals distinct boundary characteristics. As illustrated in Figures~\ref{fig:Leg_comparison} and~\ref{fig:chapadlo_results}, conventional linear post-processing strictly preserves the geometric features from raw SIMP output, while the proposed SDF-based method produces smoothed boundaries through global smoothing effects. The SDF-based approach eliminates stair-stepped interfaces while maintaining essential structural features, though this behavior may affect precision at boundary condition interfaces, requiring discretization strategies that ensure proper node positioning at predefined boundary surfaces.

\section{Conclusion}\label{sect:conclusion}
\noindent This paper presents a robust post-processing methodology that addresses the key challenges for effective topology optimization post-processing established in the introduction: geometric fidelity, volume preservation, boundary quality, mesh independence, and multi-platform compatibility. By utilizing a signed distance function approach combined with radial basis function smoothing, our method transforms element-wise density distributions from SIMP-based optimization into precisely defined geometric models with continuous boundaries. The methodology bridges the gap between raw optimization results and practical engineering applications by eliminating discretization artifacts while preserving essential structural features and maintaining computational tractability across diverse mesh configurations.

The validation across diverse test cases demonstrates three fundamental capabilities of the proposed methodology. First, the method generates exceptionally smooth geometric boundaries that preserve essential structural features, particularly in regions away from domain boundaries where sharp features are absent. This smoothing capability directly addresses stress concentration issues inherent in conventional post-processing approaches, creating continuous geometric transitions that enhance structural integrity. Second, the SDF-based framework provides precise mathematical control over volume fraction preservation through its zero-level representation, ensuring design constraint satisfaction throughout the post-processing workflow. Third, the method's ability to generate geometries that appear to originate from much finer discretizations is particularly noteworthy - while maintaining computational efficiency of the original coarse mesh optimization. This characteristic allows for achieving high-quality geometric representations without incurring the computational cost typically associated with fine-mesh topology optimization. Still, it should be noted that optimization on genuinely finer meshes would potentially reveal additional structural features such as thin membranes and ribs.

The methodology's practical value extends beyond geometric smoothing to encompass manufacturing compatibility and structural analysis requirements. The implicit boundary representation enables direct export to standard manufacturing formats without intermediate reconstruction steps, streamlining the design-to-production workflow. The integration with isosurface stuffing algorithms generates high-quality tetrahedral meshes with guaranteed dihedral angle bounds, ensuring robust numerical analysis of post-processed geometries. On the other hand, the method's global smoothing characteristics require careful consideration at boundary condition interfaces, where specialized discretization strategies may be needed to ensure proper node positioning at predefined boundary surfaces. This consideration is balanced by the method's demonstrated ability to maintain structural performance while significantly improving geometric quality, making it particularly suitable for applications where manufacturing constraints and structural optimization must be simultaneously addressed.

Future research directions could focus on incorporating manufacturing-specific constraints during the smoothing process. A particularly promising direction involves coupling the proposed post-processing method with subsequent level-set based topology optimization to achieve more efficient geometries. This combined approach would leverage smooth boundary representations obtained from our method as initial geometries for further shape refinement, potentially leading to superior structural performance. Additional opportunities exist in implementing mid-range RBFs to replace global RBFs, thereby improving computational efficiency. The current implementation, however, already provides a robust foundation for practical engineering applications where high-quality geometric representations are essential.

\paragraph{Statement:} During the preparation of this work the authors used Claude (Anthropic's AI assistant) in order to improve writing clarity and ensure consistent academic style. After using this tool/service, the authors reviewed and edited the content as needed and take full responsibility for the content of the published article.

\section*{Acknowledgements}
\noindent This research was funded by the European Union under the project Metamaterials for thermally stressed machine components (reg. no. CZ.02.01.01/00/23\_020/0008501) and the institutional support RVO:61\-38\-89\-98.\\

\noindent This work also received support from the Grant Agency of the Czech Technical University in Prague, under grant \\ No. SGS24/123/OHK2/3T/12.

\newpage

\appendix
\section{Distance function construction methodology} \label{app:distance_function}
\noindent This appendix details the practical implementation of the DF construction methodology introduced in Section~\ref{subsub:distance_function_formulation}. We present detailed implementation strategies for optimal grid resolution selection, boundary detection algorithms, and specific distance computation methods for each boundary type.

\subsection{Cartesian grid foundation} \label{subapp:grid_foundation}
\noindent The construction of an accurate SDF relies on a properly configured Cartesian grid structure. This regular grid serves as the spatial framework upon which the SDF is calculated and directly influences the quality of geometric representation. Grid spacing selection involves balancing competing requirements: coarse spacing leads to geometric detail loss, while excessively fine spacing increases computational overhead and potentially undermines the intended surface smoothing effect. This trade-off is clearly demonstrated in Figure~\ref{fig:comprehensive_sphere_analysis}, where varying grid resolutions produce distinctly different geometric representations of the same underlying topology. The following section detail the key considerations for optimal grid implementation.

\subsubsection{Optimal grid step size selection} \label{subsubapp:grid_step_size}
\noindent The optimal grid spacing is related to the size of elements in the topology optimization (TO) domain. However, this relationship cannot be expressed explicitly due to potentially arbitrary discretization schemes. While the optimal spacing can be determined iteratively by minimizing the objective function defined in the TO problem, our methodology indicates that for "reasonably" discretized TO domains where elements are roughly the same size, the shortest element edge length provides an appropriate grid spacing parameter that effectively captures topological features of the optimized design.

\subsubsection{Domain coverage and boundary handling} \label{subsubapp:grid-domain_coverage}
\noindent The regular grid must fully cover the topology optimization domain, with a certain overlap. The overlap size should be chosen to ensure that the grid boundary does not affect the resulting optimized geometry. For simplicity, in this paper, we use a standard overlap equivalent to the distance of three regular grid elements, which helps maintain the integrity of the optimized geometry without interference from the grid boundaries.

\subsubsection{Initialization strategy} \label{subsubapp:grid-initialization_strategy}
\noindent The distance field computation begins by initializing all grid nodes with a large negative value (e.g., -1e10). This initialization serves as a placeholder value, indicating nodes that have not yet been processed. Furthermore, this negative value signifies that these nodes are outside the body. The large number facilitates the comparison and updating process during computation, as it can be easily replaced with the actual distance once determined. This approach is important for simplifying the algorithm's implementation and detecting potential errors.

\subsection{Boundary classification} \label{subapp:boundary_classes}
\begin{figure}[h!]
    \centering
    \begin{subfigure}[b]{0.128\textwidth}
        \centering
        \includegraphics[width=\textwidth]{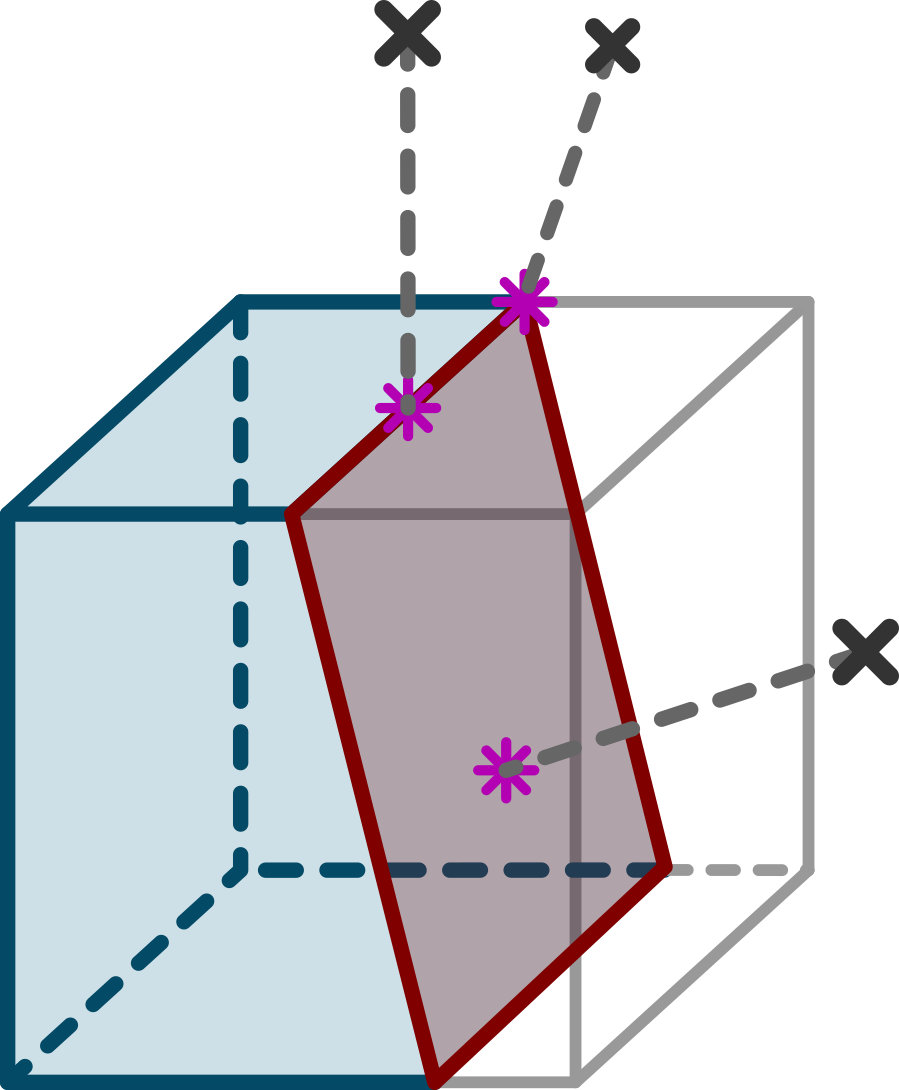}
        \caption{}
        \label{subfig:izo_projection}
    \end{subfigure}
    \hspace{0.02\textwidth} 
    \begin{subfigure}[b]{0.15\textwidth}
        \centering
        \includegraphics[width=\textwidth]{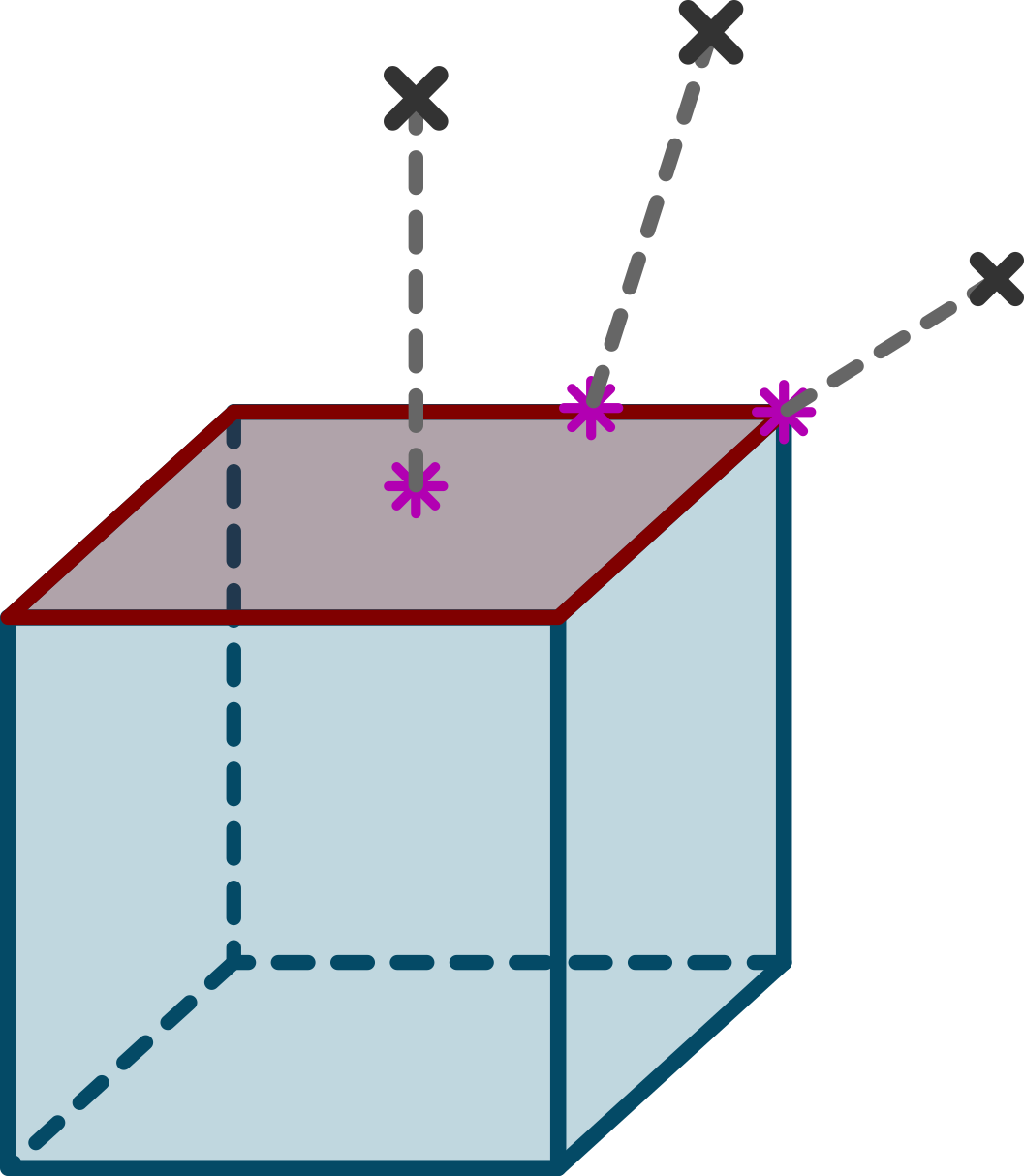}
        \caption{}
        \label{subfig:face_projection}
    \end{subfigure}
    \hspace{0.02\textwidth} 
    \begin{subfigure}[b]{0.118\textwidth}
        \centering
        \includegraphics[width=\textwidth]{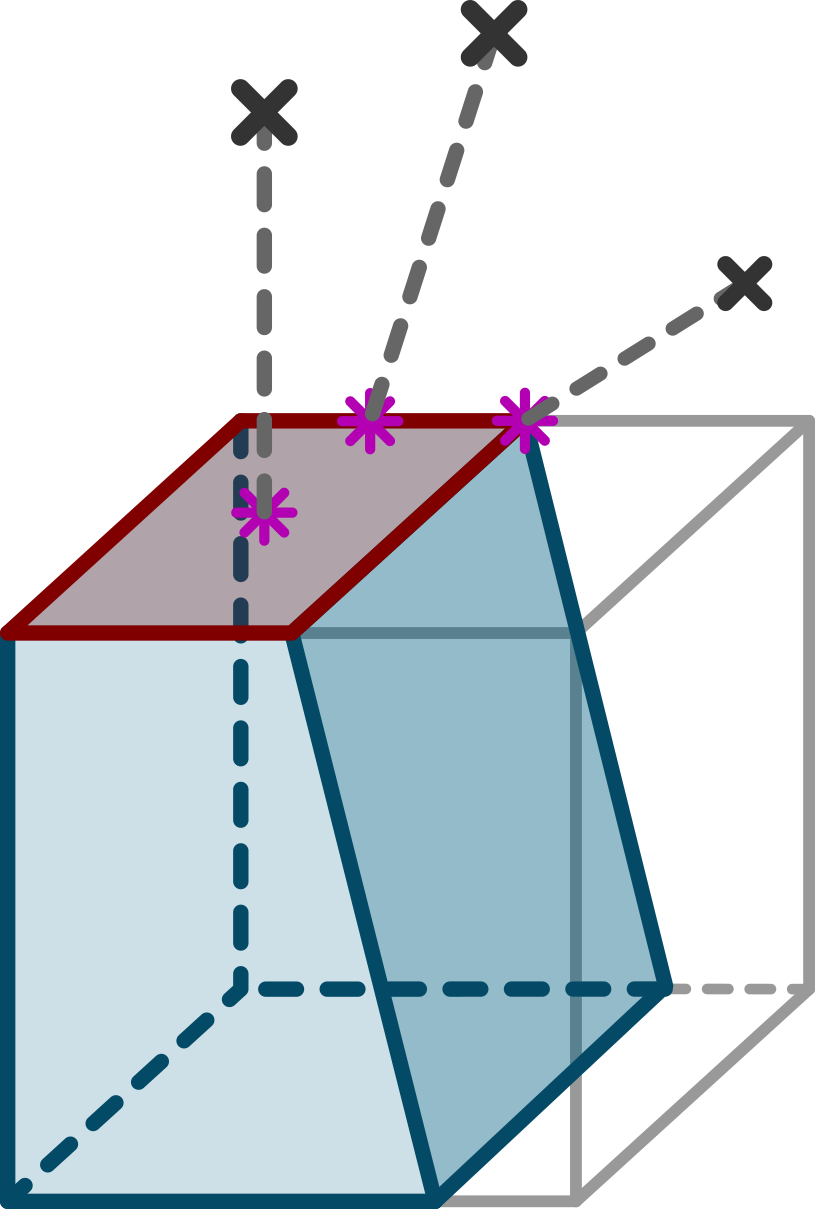}
        \caption{}
        \label{subfig:face_projection2}
    \end{subfigure}
        \caption{Demonstration of three different types of point projections onto the solid part of the element. (a) Projection on object boundary defined by isocountour. (b) Projection on the external face of the solid element. (c) Projection on the external face of a partly solid element.}
    \label{fig:projection}
\end{figure}

\noindent The DF formulation presented in Section~\ref{subsub:distance_function_formulation} requires different implementation approaches depending on the boundary type encountered. These boundary classifications emerge from analyzing both the nodal density values within elements and whether the element contains faces that form part of the topology optimization domain boundary. This systematic analysis identifies three fundamental boundary types, each requiring a specific implementation of the DF: 

\begin{enumerate}
\item \textbf{Boundary as isocontour}: All partially solid elements contain an isocontour. A DF can be constructed to this isocontour, as indicated in Figure~\ref{subfig:izo_projection}. The shortest distance can be:
\begin{enumerate}[label=\alph*)]
  \item the perpendicular projection onto the isocontour,
  \item the perpendicular projection onto the edge of the isocontour, or
  \item to the vertex of the isocontour.
\end{enumerate}
   The DF construction to the isocontour is detailed in Section \ref{subsubapp:dist2iso}.

\item \textbf{Boundary as a face of solid element}: If an element is fully solid and also lies on the boundary of the TO domain, it has at least one external face to which a DF can be constructed, as shown in Figure~\ref{subfig:face_projection}. The shortest distance can be:
\begin{enumerate}[label=\alph*)]
   \item the perpendicular projection onto the face,
   \item the perpendicular projection onto the edge of the face, or
   \item to the vertex of the face.
\end{enumerate}
   The DF construction to fully solid element faces is detailed in Section \ref{subsubapp:dist2face}.

\item \textbf{Boundary as a face of a transitional element}: The so-called transitional elements are those that are partially solid and located on the boundary of the TO domain. These elements may have external faces that are fully or partially solid. A DF can be constructed for these faces as shown in Figure~\ref{subfig:face_projection2}. The shortest distance in this case can be:
\begin{enumerate}[label=\alph*)]
  \item the perpendicular projection onto the fully filled part of the face,
  \item the perpendicular projection onto the edges of the fully filled part of the face, or
  \item to the vertex of the fully filled part of the face.
\end{enumerate}
   The DF construction to transitional element faces is detailed in Section \ref{subsubapp:dist2transface}.
\end{enumerate}

The following sections provide a detailed description of how the DF is constructed for each type of boundary.

\subsection{Distance computation methods} \label{subapp:distance_methods}
\subsubsection{Distance computation for isocontours} \label{subsubapp:dist2iso}
\noindent Each element that is partially filled contains an isocontour, which is implicitly defined by nodal densities and finite element shape functions. The DF, in this case, is constructed such that for each node of the Cartesian regular grid, the shortest distance to the isocontour is sought.

Specifically, we are looking for the isoparametric coordinates $\boldsymbol{\xi}$ of a projected point inside the element, whose physical coordinates $\boldsymbol{x}(\boldsymbol{\xi})$ are at the shortest distance $d(\boldsymbol{\xi})$ from the point of the structured grid $\boldsymbol{x}_g$ and simultaneously lie on the isocontour $\rho(\boldsymbol{\xi}) = \rho_{\mathrm{t}}$. The process of finding the local coordinates of the projected point on the density isocontour involves minimizing the distance with constraint conditions that define the isocontour value and restrict the solution to the points within the element. This can be expressed by equation \ref{eq:dist_mini}.

By restricting the solution to points located within the element, this formulation accounts for all cases of projection onto the isocontour shown in Figure~\ref{subfig:izo_projection}.

The formulation in equation \ref{eq:dist_mini} represents a nonlinearly constrained box-constrained optimization problem, requiring advanced mathematical methods for its solution. For this purpose, we utilize the NLopt library~\cite{NLopt}, specifically implementing the Sequential Least Squares Programming (SLSQP) algorithm~\cite{kraft1988software} for nonlinear programming with nonlinear constraints and bound constraints. NLopt provides an extensive collection of nonlinear optimization algorithms, with SLSQP being particularly well-suited for solving our constrained optimization problem due to its efficient handling of both equality and inequality constraints. An important computational advantage of this formulation is that the optimization problem is easily parallelizable, as each grid node can be processed independently, enabling significant performance improvements for large-scale applications.

The SLSQP algorithm's key strength lies in its efficient handling of both equality and inequality constraints through sequential quadratic programming. By creating and solving a sequence of quadratic approximations, it effectively manages the nonlinear objective function while maintaining geometric constraints for isocontour projection. This approach provides an efficient and reliable solution method for our specific isocontour projection problem.

\subsubsection{Distance computation for solid faces} \label{subsubapp:dist2face}
\noindent In the case where an element is fully filled and lies on the boundary of the TO domain, it contains at least one exterior face (the interface between the TO domain and empty space). From an implementation perspective, each exterior face is always part of only one element. The DF, in this case, is constructed such that for each node of the regular grid, the shortest distance to the external face is sought.

\begin{figure}[h!]
    \centering
    \begin{subfigure}[b]{0.145\textwidth}
        \centering
        \includegraphics[width=\textwidth]{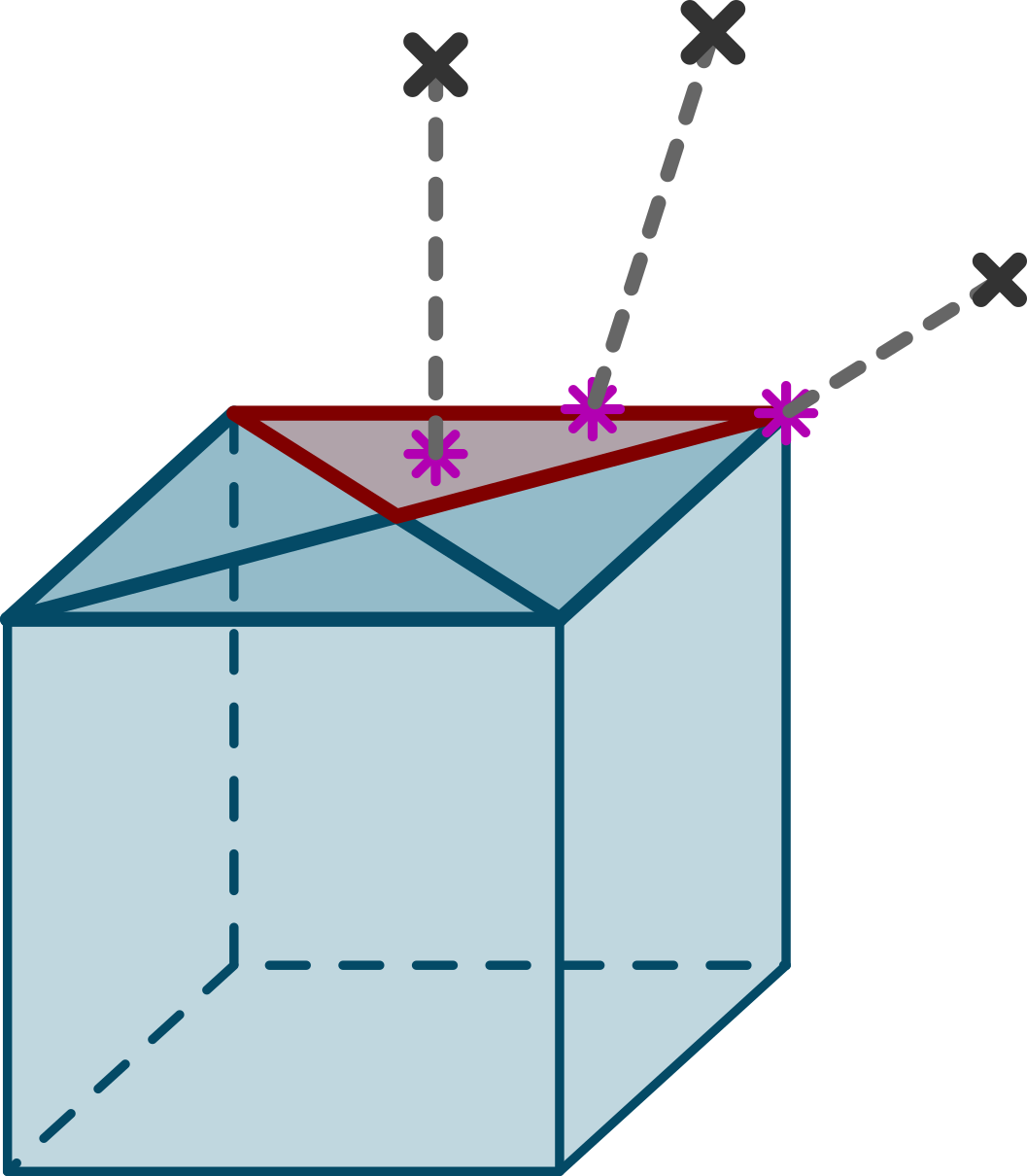}
        \caption{}
        \label{subfig:projekce_plny_tri_face}
    \end{subfigure}
    \hspace{0.02\textwidth} 
    \begin{subfigure}[b]{0.17\textwidth}
        \centering
        \includegraphics[width=\textwidth]{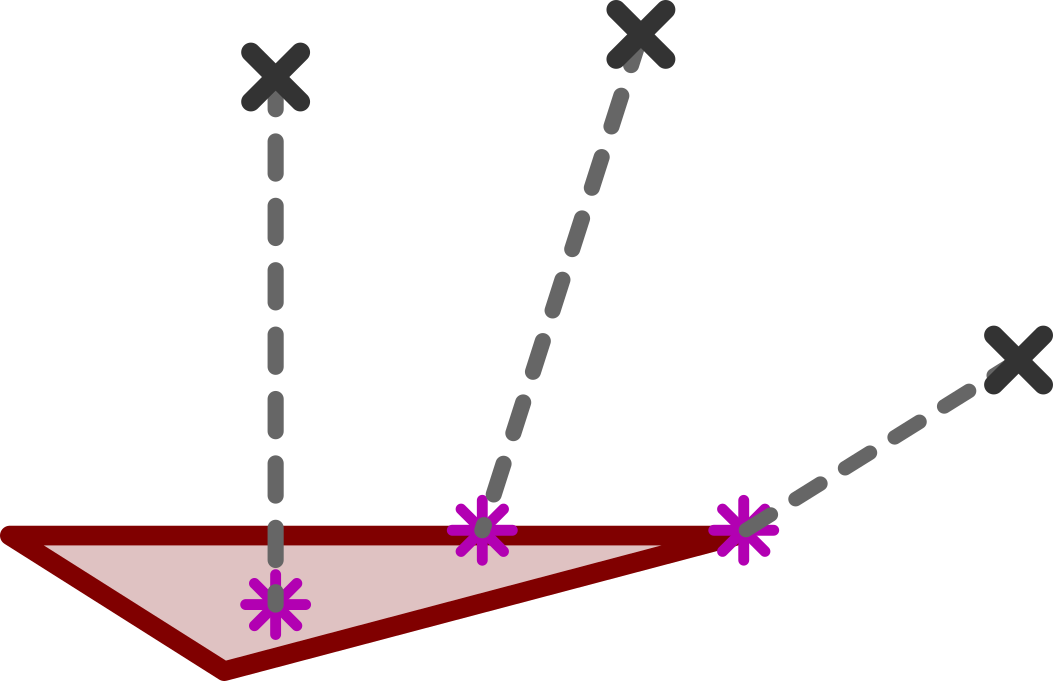}
        \caption{}
        \label{subfig:tri_face}
    \end{subfigure}
    \caption{Projection on external face of the solid element. (a) External face is descretized by triangles. (b) Problem simplification - projection on single triangle.}
    \label{fig:projekce_tri}
\end{figure}

To simplify DF calculation, it is useful to discretize these exterior faces using triangles (see Figure~\ref{subfig:projekce_plny_tri_face}). Each face is discretized into four triangles, sharing a common node at the geometric center of the face. This approach reduces the entire DF calculation problem to computing the distance between a node of the regular grid and a triangle (see Figure~\ref{subfig:tri_face}).
The shortest distance to a triangle can be:
\begin{enumerate}
\item the perpendicular projection onto the face,
\item the perpendicular projection onto the edge, or
\item distance to the triangle vertex.
\end{enumerate}

To verify whether the minimum distance is a perpendicular projection onto the triangle, barycentric coordinates $\boldsymbol{\lambda}$ are used. If the following condition is satisfied, the perpendicular projection of the grid node is inside the triangle:
\begin{equation}
\min(\boldsymbol{\lambda}) \geq 0
\end{equation}
then the coordinates of the projected point $x_p$ can be determined as follows:
\begin{equation}
  x_p = \lambda_1 x_1 + \lambda_2 x_2 + \lambda_3 x_3
\end{equation}
where $\lambda_{1,2,3}$ are the barycentric coordinates and $x_{1,2,3}$ are the coordinates of the triangle's vertices.

If the perpendicular projection is outside the triangle's face, it is necessary to check for a perpendicular projection onto one of the triangle's edges. This can be done simply using the dot product of the vector (from a triangle vertex to the grid node) and the vector describing the triangle's edge. If the perpendicular projection onto any edge of the triangle does not exist, the minimum distance to the triangle's vertices is calculated.

\subsubsection{Distance computation for transitional faces} \label{subsubapp:dist2transface}
\noindent Transition elements are those that contain an isocontour and must have at least one exterior face. The algorithm for finding projections onto the exterior face, and thus the shortest distances, is identical to the algorithm described in Section \ref{subsubapp:dist2face}. However, for this type of element, it is additionally necessary to verify that the element is solid at the point of projection onto the face.

Specifically, it must be verified that the density $\rho(\boldsymbol{\xi})$ at the point $\boldsymbol{x}_p$ is greater than or equal to the threshold density $\rho_t$. To perform this verification, we first need to look for the isoparametric coordinates $\boldsymbol{\xi}$ of a projected point. The finding of isoparametric coordinates involves minimizing the distance between the projected point $\boldsymbol{x}_p$ and the point $\boldsymbol{x}(\boldsymbol{\xi})$.

Knowing the local coordinates of the projected point $\boldsymbol{x}_p$, the density at this point can be calculated by solving equation \ref{eq:int_density}.\\
If the resulting density $\rho(\xi)$ is greater than or equal to the threshold density $\rho_t$, the point has been projected onto the full part of the element, and the results can be used for the construction of the DF. Otherwise, the calculated distance for this point is not included in the DF.

\begin{figure}[h!]
    \centering
    \begin{subfigure}[b]{0.145\textwidth}
        \centering
        \includegraphics[width=\textwidth]{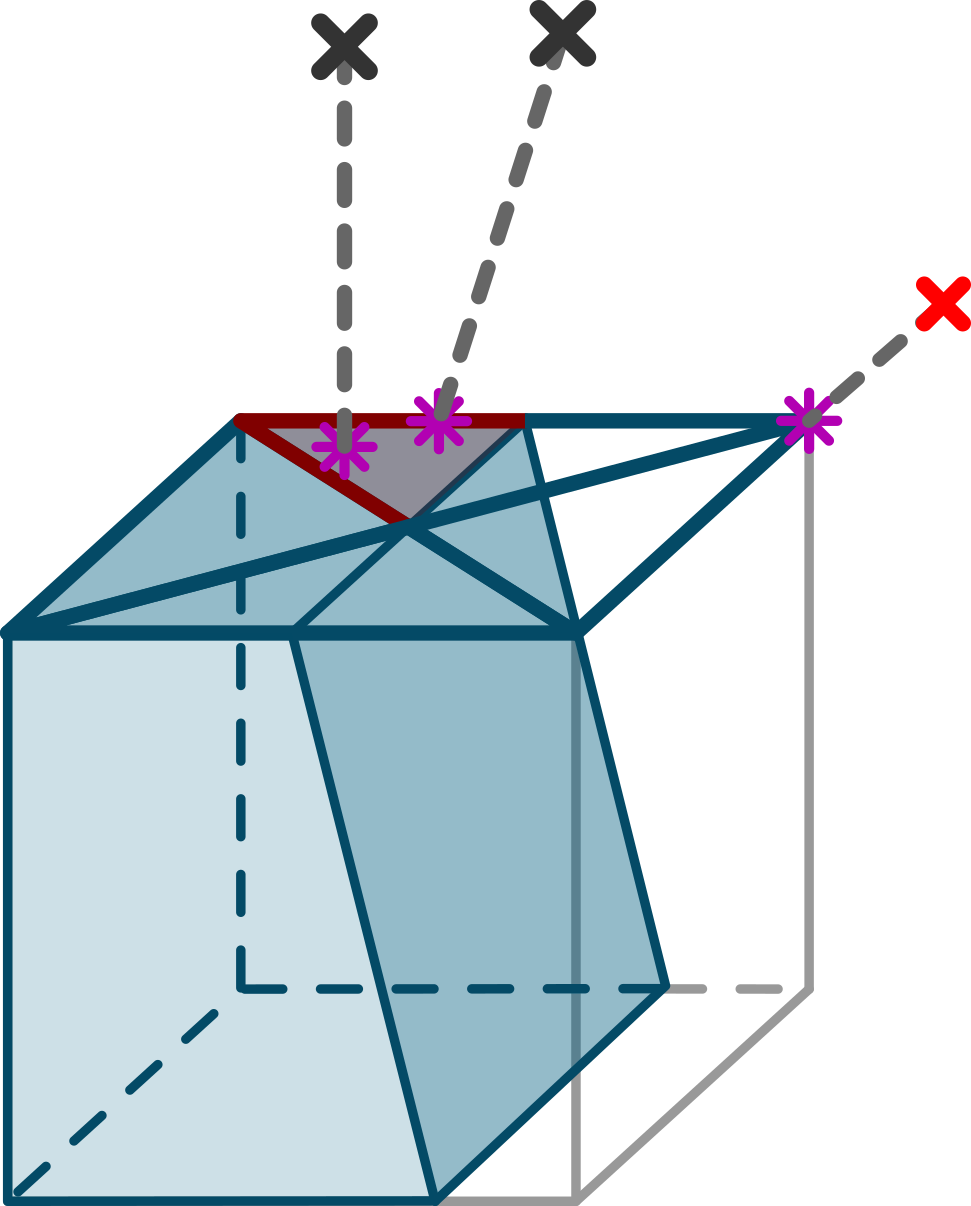}
        \caption{}
        \label{subfig:projekce_poloplny_tri_face}
    \end{subfigure}
    \hspace{0.02\textwidth} 
    \begin{subfigure}[b]{0.17\textwidth}
        \centering
        \includegraphics[width=\textwidth]{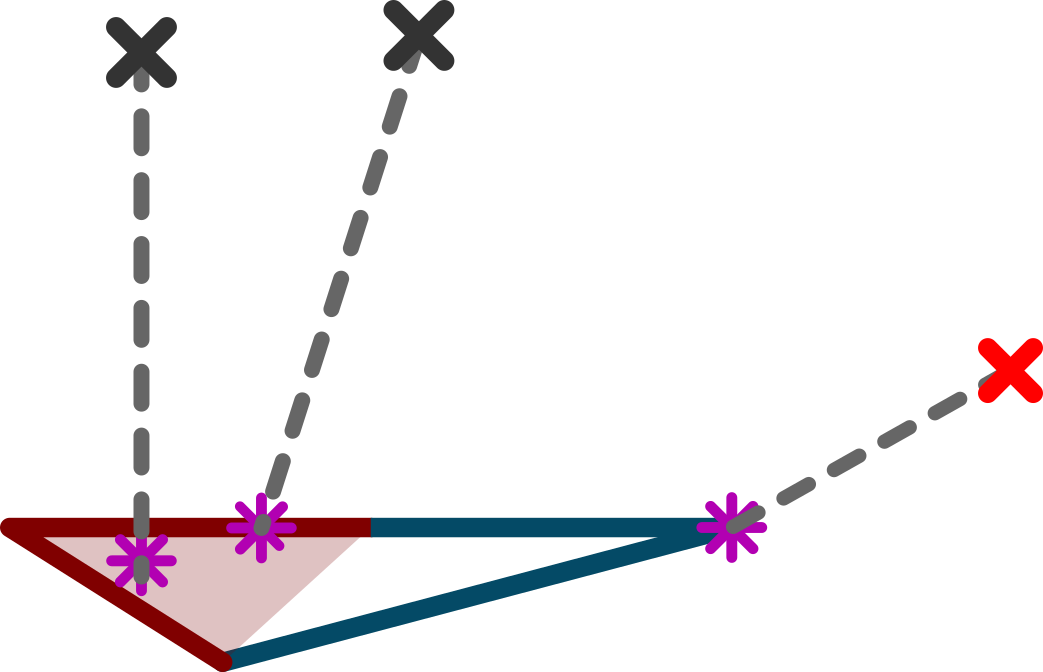}
        \caption{}
        \label{subfig:poloplny_tri_face}
    \end{subfigure}
    \caption{Projection onto the external face of the partially solid element. Black nodes are projected onto the solid part of the element, while the red node is projected onto the empty part of the element, and thus its distance cannot be used for updating the DF. (a) The external face is discretized into triangles. (b) Problem simplification - projection onto a single triangle.}
    \label{fig:poloplny_projekce_tri}
\end{figure}

\subsection{Implementation optimization strategies} \label{subapp:implementation_opt}
\subsubsection{Spatial partitioning} \label{subsubapp:AABB}
\noindent Spatial partitioning significantly reduces computational complexity by limiting distance calculations to relevant regions. In our implementation, the algorithm systematically examines each element within the topology optimization domain to identify boundary-containing elements. For each identified element, an AABB is constructed and subsequently expanded symmetrically along principal axes by a distance equal to two element edges of the regular grid. This expanded AABB creates a localized computational domain where distance function values are calculated precisely. By restricting distance computations to these expanded AABBs rather than the entire domain, we achieve substantial performance improvements while maintaining solution accuracy.

\subsubsection{Data structure optimization} \label{subsubapp:speeding}
\noindent Memory management and data organization play crucial roles in algorithm efficiency. Our implementation employs a linked list data structure to accelerate SDF construction by efficiently mapping and traversing the expanded AABBs that contain elements with boundaries. This approach allows for efficient node organization into spatially coherent lists, facilitating quick access to relevant grid points during distance calculations. As demonstrated by Fujun et al.~\cite{Fujun2000}, this approach substantially reduces execution time, particularly beneficial for large-scale problems with complex geometric representations.

\section{Data and code availability}
\noindent The implementation of the proposed SDF-based post-processing methodology, including all test cases and validation examples presented in this paper, is publicly available in the following repository:

\noindent \textbf{Repository:} \url{https://github.com/kopacja/rho2sdf.jl}

\noindent \textbf{Version:} v0.1.0\\ 

\noindent The repository contains:
\begin{itemize}
\item Complete source code for the SDF construction algorithm
\item Implementation of RBF-based geometry smoothing
\item All numerical test cases (cantilever beam, robot gripper, sphere validation)
\item Input files and mesh data for reproducing results
\end{itemize}
The code is released under the MIT License to facilitate reproducibility and encourage further research in topology optimization post-processing methods.

\printcredits

\bibliographystyle{cas-model2-names}

\bibliography{references.bib}




\end{document}